\documentclass[11pt,a4wide]{article}

\usepackage{jheppub}
\usepackage{atlasphysics}

\usepackage{subfigure}
\usepackage{rotating}

\usepackage{amsmath,graphicx}

\usepackage{booktabs}
\usepackage{multirow,xspace}

\graphicspath{{./figures/}}

\def\rqcd{\ensuremath{r_{\mathrm{OS/SS}}^{\mathrm{QCD}}}\xspace}

\def\rwjets{\ensuremath{r_{\mathrm{OS/SS}}^{W}}\xspace}

\def\nOS{\ensuremath{n_{\mathrm{OS}}^{\mathrm{bkg}}}\xspace}

\def\nSSall{\ensuremath{n_{\mathrm{SS}}^{\mathrm{bkg}}}\xspace}

\def\nOSSSqcd{\ensuremath{n_{\mathrm{OS-SS}}^{\mathrm{QCD}}}\xspace}
\def\nOSSSwjets{\ensuremath{n_{\mathrm{OS-SS}}^{W\mathrm{+jets}}}\xspace}
\def\nOSSSzjets{\ensuremath{n_{\mathrm{OS-SS}}^{Z\to\tau\tau}}\xspace}
\def\nOSSSother{\ensuremath{n_{\mathrm{OS-SS}}^{\mathrm{other}}}\xspace}
\def\meff{\ensuremath{m_{\tau\tau}^{\mathrm{eff}}}\,}

\def\nSSall{\ensuremath{n_{\mathrm{SS}}^{\mathrm{all}}}}

\def\nSSall{\ensuremath{n_{\mathrm{SS}}^{\mathrm{all}}}}

\newcommand{\mT}{\ensuremath{m_{\rm T}}}

\def\Ztoll{\ensuremath{Z/\gamma^{*} \to \ell\ell}\,}
\def\Ztoee{\ensuremath{Z/\gamma^{*} \to ee}\,}
\def\Ztomumu{\ensuremath{Z/\gamma^{*} \to \mu\mu}\,}
\def\Ztotautau{\ensuremath{Z/\gamma^{*} \to \tau\tau}\,}

\def\MET{\ensuremath{E_{\mathrm{T}}^{\mathrm{miss}}}\,}

\def\pt{\ensuremath{p_{\mathrm{T}}}\,}
\def\Wjets{\ensuremath{W+\mathrm{jets}}\,}

\def\ttbar{\ensuremath{t\bar{t}}\,}
\def\nOS{\ensuremath{n_{\mathrm{OS}}^{\mathrm{bkg}}}\,}

\def\thad{\ensuremath{\tau_{\mathrm{ had}}}}
\def\tlep{\ensuremath{\tau_{\mathrm{ lep}}}}

\def\thadp{\ensuremath{\tau^+_{\mathrm{ had}}}}
\def\thadm{\ensuremath{\tau^-_{\mathrm{ had}}}}
\def\tlepp{\ensuremath{\tau^+_{\mathrm{ lep}}}}
\def\tlepm{\ensuremath{\tau^-_{\mathrm{ lep}}}}

\def\Httll{\ensuremath{H\to\tlep\tlep}\,}
\def\Httlh{\ensuremath{H\to\tlep\thad}\,}
\def\Htthh{\ensuremath{H\to\thad\thad}\,}

\def\Httllc{\ensuremath{H\to\tlepp\tlepm}\,}
\def\Httlhc{\ensuremath{H\to\tlepp\thadm}\,}
\def\Htthhc{\ensuremath{H\to\thadp\thadm}\,}

\usepackage{preprintcover}  
\PreprintCoverPaperTitle{
\begin{boldmath}
\textbf{
Search for the Standard Model Higgs boson in the $\boldsymbol{H\to \tau^+\tau^-}$ decay mode 
in $\boldsymbol{\sqrt{s}=7 }$ \TeV{} $pp$ collisions with ATLAS
}
\end{boldmath}
}  
\PreprintIdNumber{CERN-PH-EP-2012-140}  
\PreprintCoverAbstract{
A search for the Standard Model Higgs boson decaying into a pair of $\tau$ leptons is reported. The analysis is based on a data sample of proton-proton collisions collected by the ATLAS experiment 
at the LHC and corresponding to an integrated luminosity of $4.7\,\ifb$. No significant excess over the expected background is observed in the Higgs boson mass range of 100--150\,\gev. 
The observed (expected) upper limits on the cross section times the branching ratio for $H \to \tau^+ \tau^-$ are found to be between 2.9 (3.4) and 11.7 (8.2) times the Standard Model 
prediction for this mass range.
}  
\PreprintJournalName{JHEP}  

\begin{document}


\title{
\begin{boldmath}
\textbf{
Search for the Standard Model Higgs boson in the $\boldsymbol{H\to \tau^+\tau^-}$ decay mode 
in $\boldsymbol{\sqrt{s}=7 }$ \TeV{} $pp$ collisions with ATLAS
}
\end{boldmath}
}
\author{The ATLAS Collaboration}

\abstract{
A search for the Standard Model Higgs boson decaying into a pair of $\tau$ leptons is reported. The analysis is based on a data sample of proton-proton collisions collected by the ATLAS experiment 
at the LHC and corresponding to an integrated luminosity of $4.7\,\ifb$. No significant excess over the expected background is observed in the Higgs boson mass range of 100--150\,\gev. 
The observed (expected) upper limits on the cross section times the branching ratio for $H \to \tau^+ \tau^-$ are found to be between 2.9 (3.4) and 11.7 (8.2) times the Standard Model 
prediction for this mass range.
}

\maketitle


\section{Introduction}
\label{sec:intro}
The Higgs boson is the only fundamental particle in the Standard Model (SM) of particle 
physics that has not yet been observed. It is predicted by the Higgs mechanism~\cite{prl_13_321,pl_12_132,prl_13_508,Guralnik:1964eu,pr_145_1156,Kibble:1967sv},
which in the SM gives mass to particles. 
The search for the Higgs boson is a centrepiece of the Large Hadron Collider (LHC) physics programme.

An indirect constraint on the Higgs boson mass of $m_H < 185\GeV$ at the $95\%$ confidence 
level (CL) has been set using global fits to electroweak precision data~\cite{lepew:2010vi}. Direct searches at LEP and the Tevatron have placed exclusion limits at  95\% CL 
for $m_H<114.4\GeV$ and in the region $147\GeV < m_H < 179\GeV$~\cite{Barate:2003sz,TEVNPH:2012ab}, respectively.
The results of searches in various channels using data corresponding to an integrated luminosity of up to 
$5\, \ifb$ have recently been reported by both the ATLAS and CMS 
Collaborations~\cite{atlashiggspaper,Chatrchyan:2012tx} excluding the mass range between $112.9\GeV$ and $115.5\GeV$ and the region between $127\GeV$ and $600\GeV$, again at the 95\% CL.

In the Higgs boson mass range 100-150~GeV, the $H\to \tau^+\tau^-$ decay mode is a promising 
channel for the search at the LHC with 
branching ratios between 8\% and 1.8\%. The $H\to \tau^+\tau^-$ search is complementary to searches 
with other decays in the same mass range and enhances the overall sensitivity. 
Also, if the Higgs boson is discovered, the measurement of the $H\to \tau^+\tau^-$ decay rate 
provides a test of the SM prediction for the $\tau$ Yukawa coupling.

The process with the largest cross section to produce a SM Higgs boson at the LHC is 
gluon fusion $gg\to H$. However, Higgs boson production via vector boson ($WW$, $ZZ$) fusion 
$qq\to qqH$ (VBF) and via Higgs-strahlung $qq\to VH$, in association with a hadronically 
decaying vector boson ($V=W$ or $Z$), are highly relevant as well because they lead to additional 
jets in the final state, which provide distinct experimental signatures. In particular 
the VBF topology of two high-energy jets with a large rapidity separation offers a good 
discrimination against background processes. 
For all production processes above, the Higgs boson is
typically more boosted in the transverse plane if there are additional high-\pt jets in the event, 
which increases the transverse momentum 
of the $\tau$ decay products and thus facilitates the measurement of the $\tau^+\tau^-$ 
invariant mass and the discrimination of the signal from background processes. 

This paper presents SM Higgs boson searches in the the \Httllc, \Httlhc, and \Htthhc\ channels,\footnote{
Charge-conjugated decay modes are implied.
Throughout the remainder of the paper, a simplified notation without the particle charges is used. 
} where
$\tlep$ and $\thad$ denote leptonically and hadronically decaying $\tau$ leptons, respectively.
The data analyses use proton-proton ($pp$) collisions at $\sqrt{s} = 7~\mbox{TeV}$ collected by 
the ATLAS experiment in 2011, which correspond to an integrated luminosity of 4.7 fb$^{-1}$.
In order to enhance the sensitivity of the search, the selected events are analysed in several separate 
categories according to the number and topology of reconstructed jets.

\section{Data and Monte Carlo simulated samples}
\label{sec:samples}

The ATLAS detector is  a multipurpose apparatus with a forward-backward
symmetric cylindrical geometry and nearly 4$\pi$ coverage in solid
angle~\cite{atlas-det}. It consists of an inner tracking detector surrounded by a thin
superconducting solenoid, electromagnetic and hadronic calorimeters, and an
external muon spectrometer incorporating three large superconducting air-core toroid magnets.
Electrons,
muons, $\tau$ leptons and jets can be reconstructed and identified in the ATLAS 
detector.\footnote{
  ATLAS uses a right-handed coordinate system with
  its origin at the nominal interaction point (IP) in the centre of the detector
  and the $z$-axis along the beam pipe. The $x$-axis points from the IP to the centre of
  the LHC ring, and the $y$-axis points upwards. Cylindrical coordinates $(r,\phi)$ are used
  in the transverse plane, $\phi$ being the azimuthal angle around the beam pipe.
  The pseudorapidity is defined in terms of the polar angle $\theta$ as
  $\eta=-\ln\tan(\theta/2)$.}
Only data taken with all sub-systems relevant to this analysis operational are used. 
This results in an integrated luminosity of 4.7 fb$^{-1}$ for the full 2011 data sample.

The signal contributions considered here include
the gluon fusion production process, 
the VBF production process,
and the  Higgs-strahlung $VH$ 
production in association with a hadronically-decaying vector boson $V=W,Z$. 
For the decay of the Higgs boson, the $H\rightarrow\tau\tau$ mode is considered. 
Next-to-next-to-leading-order (NNLO) Quantum Chromodynamics (QCD) corrections, soft-gluon resummations calculated in the next-to-next-to-leading-log-approximation and next-to-leading-order  (NLO) 
electroweak (EW) 
corrections are applied to the signal cross sections 
for the
$gg\to H$ production process~\cite{Djouadi1991,Dawson1991,Spira1995,Harlander:2002wh,Anastasiou:2002yz,Ravindran:2003um,Catani:2003zt,Aglietti:2004nj,Actis:2008ug,Anastasiou:2008tj,deFlorian:2009hc,Baglio:2010ae}. The cross sections of the VBF process are calculated with full NLO QCD and electroweak corrections~\cite{Ciccolini:2007jr,Ciccolini:2007ec,Arnold:2008rz}, and approximate NNLO
QCD corrections~\cite{Bolzoni:2010xr}.
The $WH$/$ZH$ processes are calculated with NNLO QCD corrections~\cite{Han:1991ia,Brein:2003wg} and NLO EW radiative corrections are applied~\cite{Ciccolini:2003jy}.
The Higgs boson decay branching ratios are calculated using
HDECAY~\cite{Djouadi:1997yw}.
The $gg\to H$ and VBF processes are modelled using the
POWHEG~\cite{Alioli:2008tz,Nason:2009ai}
Monte Carlo (MC) generator,
interfaced to 
PYTHIA~\cite{pythia} for showering and hadronisation. In the $gg \to H$ process, the Higgs boson $\pt$
spectrum is reweighted to agree with the prediction of the HqT program~\cite{deFlorian:2011xf}. 
The associated $VH$ production process is modelled using PYTHIA.

ALPGEN~\cite{alpgen}, interfaced to HERWIG~\cite{herwig}, with 
the MLM matching scheme~\cite{alwall-2008-53} is used to model the production
of single $W$
and
$Z/\gamma^{\ast}$
bosons decaying to charged leptons
 in association with jets.
MC@NLO~\cite{mcatnlo} is used to model $t\bar{t}$, $WW$, $WZ$ and $ZZ$
production processes, using HERWIG for the parton shower and hadronisation and 
JIMMY~\cite{jimmy} for the underlying event modelling.
AcerMC~\cite{Kersevan:2004yg} is used to model single top-quark production for all  
three production channels ($s$-channel, $t$-channel, and $Wt$ production).

The TAUOLA~\cite{Jadach:1993hs} and PHOTOS~\cite{Golonka:2005pn} 
programs are used to model the decay of $\tau$ leptons and the Quantum Electrodynamics (QED) radiation of photons, 
respectively.

The set of parton distribution functions (PDF) CT10~\cite{ct10} is used for the MC@NLO samples, 
as well as CTEQ6L1~\cite{Pumplin:2002vw} for the ALPGEN samples, and MRST2007~\cite{mrst} for the
PYTHIA and HERWIG samples.
Acceptances and efficiencies are based on a simulation of the
ATLAS detector using GEANT4~\cite{GEANT4, atlassim}. 
Since the data are affected by the
detector response to multiple interactions (pileup) occurring in the same or nearby
bunch crossings, the simulation includes a
treatment of the event pileup conditions present in the 2011 data.

The \Ztotautau\ background processes are modelled with a $\tau$-embedded $\Ztomumu$ data sample 
as described in Section~\ref{sec:backgrounds}.

\section{Selection and reconstruction of physics objects}
\label{sec:objects}

Electron candidates are  formed from an energy deposit in the electromagnetic calorimeter and associated to a track measured in the inner detector. They are selected if they have a transverse energy $\ET > 15\gev$, lie within
$|\eta|<2.47$ but outside the transition region between the barrel and end-cap 
calorimeters ($1.37<|\eta|<1.52$), and 
meet quality requirements based on the expected shower shape~\cite{egammapaper}.  

Muon candidates are formed from a track measured in the inner detector and linked to a track 
in the muon spectrometer~\cite{ATLASWZFirstConfNote}.  They are required 
to have a transverse momentum $\pt > 10\GeV$ and to lie within $|\eta| < 2.5$. 
Additionally, the difference 
between the $z$-position of the point of closest approach of the muon inner detector track 
to the beam-line and the $z$-coordinate of the primary vertex is required to be less 
than 1~cm.\footnote{The primary vertex is defined as the vertex with the largest $\sum p^2_T$ of the associated tracks. }  This requirement reduces the contamination due to cosmic ray muons and beam-induced 
backgrounds. 
Muon quality criteria based on, e.g., inner detector hit requirements are applied
in order to achieve a precise measurement of the muon momentum and reduce the 
misidentification rate.

Identified electrons and muons are required to be isolated: the additional transverse energy in the electromagnetic and 
hadronic calorimeters must be less than 8\% (4\%) of the electron 
transverse energy (muon transverse momentum) in a cone of radius $\Delta R= \sqrt{(\Delta \eta)^{2}+ (\Delta 
\phi)^{2}}=0.2$ around the electron (muon) direction.
The sum of the transverse momenta of all tracks  with \pt\ above $1\gev$ located within a cone of radius $\Delta R=0.4$ around the 
electron (muon) direction and originating from the same primary vertex must be less than 6\% of the 
electron transverse energy (muon transverse momentum).

Jets are reconstructed using the anti-$k_{\mathrm{t}}$ algorithm~\cite{AntiKT} with a 
distance parameter value of $R=0.4$, taking as input three-dimensional 
noise-suppressed clusters in the calorimeters. 
Reconstructed jets with $\pt > 20\GeV$ and within $|\eta| < 4.5$ are selected. Events are 
discarded if a jet is associated with out-of-time activity or calorimeter noise. 
After having associated tracks to jets by requiring $\Delta R < 0.4$ between tracks and the jet direction, 
a jet-vertex fraction (JVF) is computed for each jet as the scalar $\pt$ sum of all associated
tracks from the primary vertex divided by the scalar $\pt$ sum of all tracks associated with the jet.
Conventionally, $\mbox{JVF}=-1$ is assigned to jets with no associated tracks.
Jets with $\left|\eta\right|<2.4$ are required to have $|\mbox{JVF}|>0.75$ in order to 
suppress pileup contributions.
In the pseudorapidity range $\left|\eta\right|<2.5$, $b$-jets are identified  
using a tagging algorithm based on the discrimination power of the impact parameter 
information
and of the reconstruction of the displaced vertices of the hadron decays inside the jets~\cite{ATLAS-btag-algs}.
The $b$-tagging algorithm has an average efficiency of 58\% for $b$-jets in $t\overline{t}$ events~\cite{ATLAS-btag-eff}. The 
corresponding light-quark jet 
misidentification probability is 0.1--0.5\%, depending on the jet $\pt$ and $\eta$~\cite{ATLAS-btag-mis}.

Hadronic decays of $\tau$ leptons are characterised by the 
presence of one or three charged hadrons accompanied by a neutrino and possibly neutral hadrons, 
which results in a collimated shower profile in the calorimeters and only a few nearby 
tracks.
The visible decay products are combined into $\thad$ candidates. 
These candidates are reconstructed as jets, which are re-calibrated to account for the
different calorimeter response to hadronic  decays as compared to hadronic jets.
The four-momentum of the $\thad$ candidates are reconstructed from the energy deposits in the 
calorimeters and the rejection of jets misidentified as hadronic $\tau$ decays is performed by a 
multivariate discriminator based on a boosted decision tree~\cite{TauID}
that uses both tracking and calorimeter information. The identification is optimised to be
50\% efficient while the jet misidentification probability 
is kept below 1\%.
A $\thad$ candidate must lie within $\left|\eta\right|<2.5$, have a 
transverse momentum greater than $20\,\gev$,
one or three associated tracks (with $p_{\text{T}} > 1~\mbox{GeV}$) 
and a total charge of $\pm$1 computed from the associated tracks. 
Dedicated electron and muon veto algorithms are used.

When different objects selected according to 
the above criteria overlap with each other geometrically (within 
$\Delta R < 0.2$), only one of them is 
considered for further analysis. The overlap is resolved by selecting muon, 
electron, $\thad$ and jet candidates in this order of priority.

The magnitude of the missing transverse momentum~\cite{ATLASmetNEW} ($\met$) is reconstructed  including 
contributions from 
muon tracks 
and
energy deposits in
the calorimeters. 
Calorimeter cells belonging to three-dimensional noise-suppressed clusters are used and they are
calibrated taking into account the reconstructed physics object to which they belong.

\section{Preselection}
\label{sec:preselection}

An initial selection of events is performed by  requiring a vertex from the primary $pp$ collisions that is consistent with the beam spot position, 
with at least three associated tracks, each with $\pt > 500\MeV$. Overall quality criteria are 
applied to suppress events with fake \met, produced by non-collision activity such as cosmic ray muons, 
beam-related backgrounds, or noise in the calorimeters.

The $\ell\ell$, $\ell\thad$ and $\thad\thad$ final states.\footnote{Here $\ell$ denotes an electron or a muon (also referred to below as light leptons).
}
considered in this search are defined in a mutually exclusive way: a requirement of exactly two, one, or zero electrons or muons is imposed, respectively.

\boldmath
\subsection{\Httll}  
\label{sec:presel_leplep}
\unboldmath
Signal events in this channel are selected by requiring exactly two isolated and oppositely-charged light leptons (electrons and/or muons).
Single lepton and di-lepton triggers are used to preselect the data. 
The trigger object quality requirements were tightened during
the data-taking period to cope with increasing instantaneous luminosity.
The single muon 
trigger requires $\pt>18\,\gev$;
 for the single electron trigger the $\et$ threshold changes from $20\,\gev$ 
to $22\,\gev$ depending on the LHC instantaneous luminosity; the di-muon trigger requires $\pt>15\,\gev$ for the 
leading muon and $\pt>10\,\gev$ for the sub-leading muon;  the di-electron trigger requires $\ET>12\,\gev$
for each of the two electrons; the $e\mu$ trigger requires $\ET>10\,\gev$ for the electron 
and $\pt>6\,\gev$ for the muon. 
In addition to the trigger requirements, the preselection requires $\ET > 22\GeV$ if the
electron satisfies only the single electron trigger. The $\ET$ requirement is increased to 24~\GeV\ when the trigger threshold is $22\,\gev$.  
If a muon is associated only with the single muon trigger object, it is required to have $\pt > 20\GeV$.
For the $e\mu$ channel the di-lepton invariant mass is required to be in 
the range of $30\,\gev<m_{\ell\ell}<100\,\gev$, whereas for the $ee$ and $\mu\mu$ channels 
$30\,\gev<m_{\ell\ell}<75\,\gev$ is required, reducing the contamination from $\Ztoll$.

\boldmath
\subsection{\Httlh}  
\unboldmath
Signal events in this channel are characterised by exactly one isolated light 
lepton $\ell$, a $\thad$ candidate, and large \MET\ due to the undetected 
neutrinos. 
For the  $e\thad$ ($\mu\thad$) final states, events are preselected using the single electron (muon) trigger described in Section~\ref{sec:presel_leplep}. 
Exactly one electron with $E_\mathrm{T}>25\,\gev$ or one muon with
$p_\mathrm{T}>20\,\gev$, and one oppositely-charged
$\thad$ candidate with $\pt> 20 \gev$ are required in the event. Events with
more than one electron or muon candidate are rejected to suppress events 
from
$Z/\gamma^* \to\ell^+\ell^-$  decays and from \ttbar\ or single top-quark
production.\footnote{For the purpose of vetoing  additional light leptons, the isolation
requirements are removed from the muon selection and a looser identification requirement is
used for electrons. The transverse momentum (energy) threshold for muons (electrons) is lowered to
$10\,\gev$ ($15\,\gev$).}
The transverse mass of the lepton and \MET\ is calculated as
\begin{equation}
\mT =
\sqrt{2p_{\mathrm{T}}^{\mathrm{\ell}}E_{\mathrm{T}}^{\mathrm{miss}}(1-\cos\Delta\phi)}~,
\end{equation}
where $p_{\mathrm{T}}^{\mathrm{\ell}}$ denotes the magnitude of the transverse momentum of the lepton and
$\Delta\phi$ is the angle between the lepton and \MET\ directions in the plane perpendicular to the beam
direction. In order to reduce contributions from the $W$+jets and $t\bar{t}$ background   
processes, only events with $\mT<30$~GeV are considered for further analysis.
In addition, $\met$ is used for further event selection and
categorisation, as described in Section~\ref{subsec:categories}.

\boldmath
\subsection{\Htthh}
\unboldmath
Signal events in this channel are characterised by two 
identified hadronic $\tau$ decays and large $\met$ from the undetected 
neutrinos. The corresponding event selection starts with a double hadronic $\tau$ trigger, 
where the $\pt$ 
thresholds are 29 GeV and 20 GeV for the leading and sub-leading hadronic $\tau$ objects, 
respectively. A requirement of exactly zero charged light leptons, as defined in Section~\ref{sec:objects}, is imposed.
Two identified opposite charge $\thad$ candidates with  $\pt>35\,\gev$ and $\pt>25\,\gev$ are required, each matching
a $\tau$ trigger object~\cite{atlastrigger}.

\section{Analysis categories}
\label{subsec:categories}
For further analysis, the selected event samples are split into several categories according to the number and topology of reconstructed jets.
The sensitivity of the search is usually higher for categories where the presence of one or more jets is required, as discussed in Section~\ref{sec:intro}, 
but events without any reconstructed high-\pt jets are also considered in order to maximise the sensitivity.

\boldmath
\subsection{\Httll}
\unboldmath
Four categories defined by their jet multiplicity and kinematics are used for this channel: 
$H+\!2$-jet VBF, $H+\!2$-jet $VH$, $H+\!1$-jet and $H+\!0$-jet. The first two
categories require the presence of at least two jets and the cuts are optimised in one case for the VBF process~\cite{Rainwater:1998kj,Plehn:1999xi,Asai:2004ws},
and in the other  for the $VH$ and $gg \to H$ processes~\cite{Mellado:2004tj}. 

The $H+\!0$-jet category uses an inclusive selection to collect 
part of the signal not selected by the categories with jets.
In the $H+\!0$-jet category, only the $e\mu$ final state is considered because of the overwhelming 
\Ztoll\ background in the $ee$ and $\mu\mu$ final states. In order to suppress the \ttbar\ background, it is
required that the di-lepton azimuthal opening angle be $\Delta\phi_{\ell\ell}>2.5\,$rad
and that the leptonic transverse energy be $H^{\textrm{lep}}_\mathrm{T} = p_\mathrm{T}^{\ell 1}+p_\mathrm{T}^{\ell 2}+\met< 120\,\gev$, 
where $p_\mathrm{T}^{\ell 1}$ and $p_\mathrm{T}^{\ell 2}$ are the magnitudes of the transverse momenta of the leading and sub-leading
leptons, respectively.

In categories with jets  ($H+\!2$-jet VBF, $H+\!2$-jet $VH$ and $H+\!1$-jet), the presence of a hadronic jet with a transverse momentum $\pt>40\,\gev$ is required and, to suppress the \ttbar\ 
background, the event is rejected if any jet with $\pt>25\,\gev$ is identified as a $b$-jet. 
In addition, $\met>40\,\gev$ ($\met>20\,\gev$) for the $ee,\mu\mu$ ($e\mu$) channels is also required. 

The collinear approximation technique~\cite{cll2} 
is used to reconstruct the kinematics of the $\tau\tau$ system.
The approximation is based on two 
assumptions: that the neutrinos from each $\tau$ decay are nearly collinear with the corresponding visible $\tau$ decay products and 
that the $\met$ in the event is due only to neutrinos. In this case, the total invisible momentum carried away by neutrinos 
in each $\tau$ decay can be estimated from the polar and azimuthal angles of the visible products of each $\tau$ decay. Then, the invariant mass of the $\tau\tau$ system can be calculated as  $m_{\tau\tau}={m_{\ell\ell} / \sqrt{x_1\cdot x_2} }$, where $x_1$ and $x_2$ are the momentum fractions of the 
two $\tau$ candidates carried away by their visible decay products.
Events that do not satisfy $0.1<x_1,x_2<1.0$ are rejected. In categories with jets, there is an additional requirement that
$0.5\,$rad $<\Delta\phi_{\ell\ell}<2.5\,$rad to suppress the \Ztoll\ background.

For the $H+\!2$-jet categories, 
a subleading jet with $\pt>25\,\gev$ is required in addition. For the 
$H+\!2$-jet VBF category, a pseudorapidity difference between the two selected jets of 
$\Delta\eta_{jj}>3$ and a di-jet invariant mass of $m_{jj}>350\,\gev$ are required.
Finally, the event is rejected in the $H+\!2$-jet VBF category if any additional jet with $\pt>25\,\gev$ and $\left|\eta\right|<2.4$
is found in the pseudorapidity range between the two leading jets.

For the $H+\!2$-jet $VH$ category, the requirement on the pseudorapidity separation of the jets and on the  di-jet invariant mass 
are instead: $\Delta\eta_{jj}<2$ and $50\,\gev<m_{jj}<120\,\gev$.

Only events failing the cuts for the $H+\!2$-jet categories are considered in the 
$H+\!1$-jet category.  
For the $H+\!1$-jet category, the invariant mass of the two $\tau$ leptons and the leading jet is required to fulfil
$m_{\tau\tau j}>225\,\gev$, where the $\tau$ momenta are taken from the collinear approximation.  The main Higgs production mechanism in this category is the $gg \to H$ process plus a high-$\pt$ parton.

The $m_{\tau\tau}$ calculated with the collinear approximation (``collinear mass'') is used in categories with jets. Because  this variable displays poor resolution in the $H+\!0$-jet category due to the back-to-back configuration of the two leptons, the effective mass ($\meff$), defined as the invariant mass of the two leptons and the $E_{\mathrm{T}}^{\mathrm{miss}}$, is used instead.

\boldmath
\subsection{\Httlh}
\label{subsec:lhcat}
\unboldmath
The selected data are split into seven categories based on jet properties and \MET.

The $H+\!2$-jet VBF category includes all selected events with $\MET>20\,\gev$
and at least two jets with $\pt> 25$~GeV, where the two
leading jets are found in opposite hemispheres of the detector ($\eta_{jet1}\cdot\eta_{jet2}<0$), 
with $\Delta\eta_{jj}>3$ and $m_{jj}>300\,\gev$. Both the lepton and the $\thad$ candidate are required to be found in the
pseudorapidity range between the two leading jets.
Due to the limited size of the selected event samples, the VBF category combines 
the $e\thad$ and $\mu\thad$ final states.

Two $H+\!1$-jet categories include all selected events with $\MET>20\,\gev$  and at least one 
jet with $\pt>25\,\gev$, that
fail the VBF selection. The $e\thad$ and $\mu\thad$ final states are considered 
separately.

Four $H+\!0$-jet categories include all selected events without any jet with $\pt>
25$~GeV. The $e\thad$ and $\mu\thad$ final states are considered separately. 
In addition, the analysis is separated into  events with \MET$>20$~GeV and \MET$<20$~GeV. 
The low-\MET\ region is included here because, in the absence of high-$\pt$ jets, the Higgs 
decay products, including the neutrinos, are typically less boosted than for events with additional jet activity.

For each category,
the mass of the $\tau\tau$ system is reconstructed using the  Missing Mass Calculator (MMC)~\cite{Elagin:2010aw}.
This technique provides a full reconstruction of
event kinematics in the $\tau\tau$ final state with 99\% efficiency and 13--20\%
 resolution in $m_{\tau\tau}$, depending on the event topology 
(better resolution is obtained for events with high-\pt\ jets). 
Conceptually, the MMC is a more sophisticated version of the collinear approximation. 
The main improvement comes from requiring that relative orientations of the neutrinos 
and other decay products are consistent with the mass and kinematics of a $\tau$ lepton decay. 
This is achieved by maximising a probability defined in the kinematically allowed 
phase space region.

\boldmath
\subsection{\Htthh}
\unboldmath
In the \Htthh channel, only a single $H+\!1$-jet category is 
defined. 
After selecting two hadronic $\tau$ candidates, the collinear mass approximation cuts $0<x_1,x_2<1$ are applied. Events at this stage are  used as a control sample to derive the normalisation of the
$\Ztotautau$ background.
Then, events are selected if $\MET>20\,\gev$ and if the leading jet has a transverse momentum $\pt>40\,\gev$. The two $\tau$ 
candidates are required to be separated by $\Delta R(\tau,\tau)<2.2$. 
Also, only events with an invariant mass of the $\tau\tau$ pair and the leading jet $m_{\tau\tau j}>225\,\gev$ 
are considered for further analysis. The event selection criteria described here are effective against the multi-jet and $\Ztotautau$ backgrounds.
The collinear mass approximation is used for the Higgs mass reconstruction.

\section{Background estimation and modelling}
\label{sec:backgrounds}

The background composition 
and normalisation are determined using data-driven methods and the simulated event samples 
described in Section~\ref{sec:samples}.

The main background to the Higgs boson signal in all selected final states is the largely 
irreducible \Ztotautau\ process. 
While it is not possible to select a Higgs signal-free \Ztotautau\ sample directly from the data,
this background 
is still modelled in a data-driven way, by choosing a control sample where the expected signal contamination is negligible.
\begin{figure}[t!]
  \centering
  \subfigure[\MET\ ]{
     \includegraphics[width=0.45\textwidth]{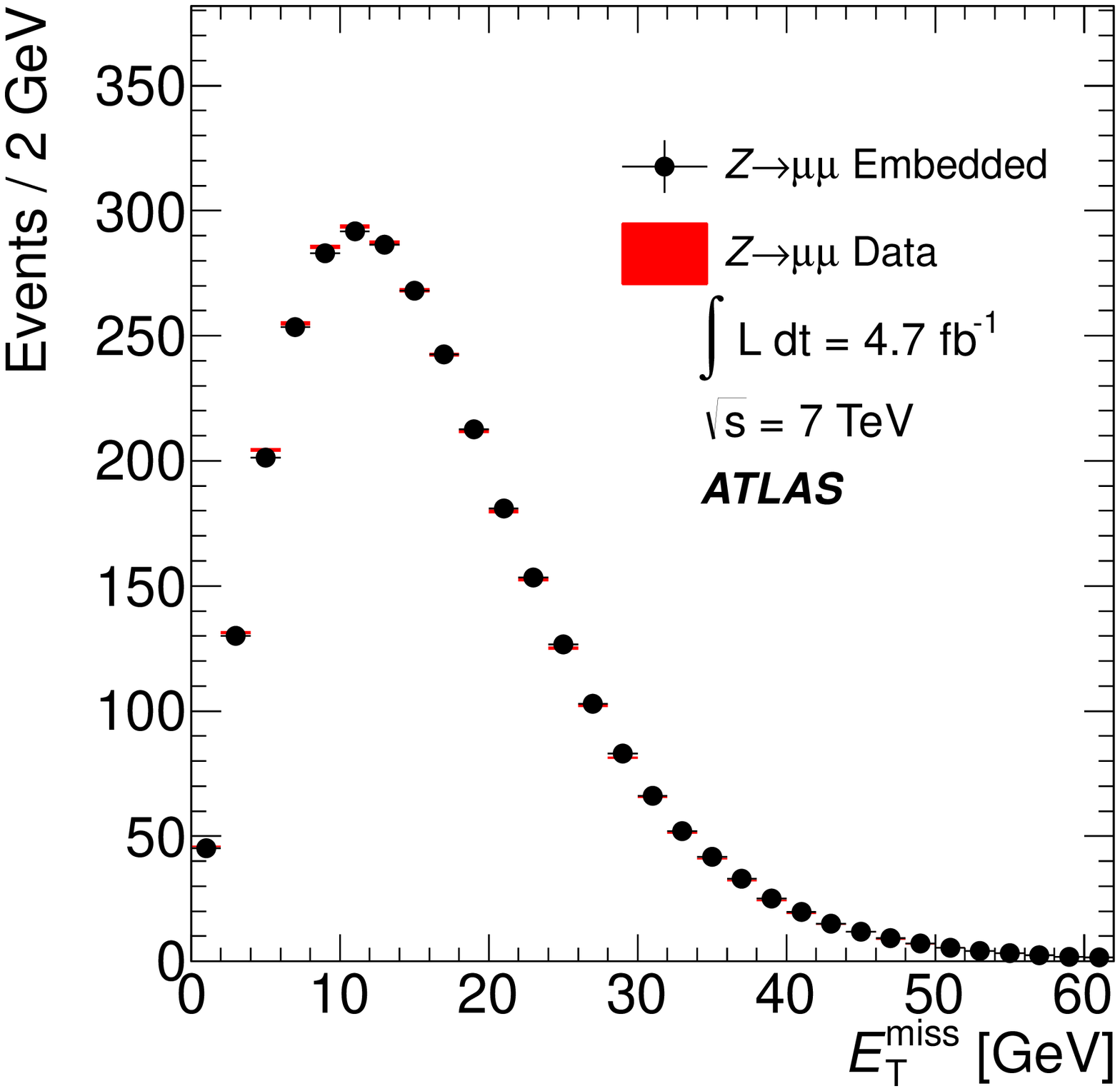}     
  }
  \subfigure[Invariant mass $m_{\tau\tau}$  ]{
     \includegraphics[width=0.45\textwidth]{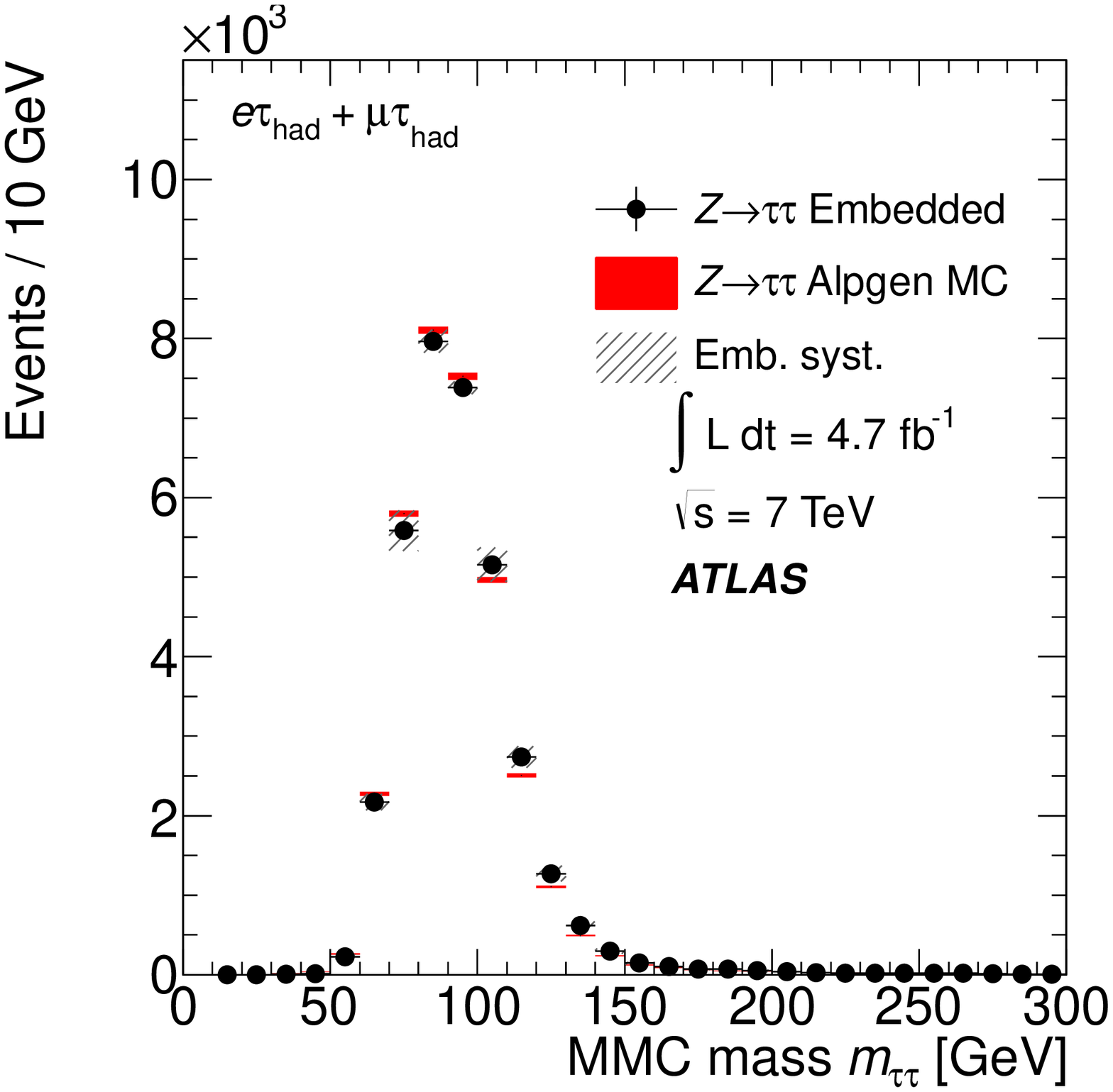}    
  }
\caption[]{(a) \MET\ distributions for the \Ztomumu\ data before and after muon embedding and (b) MMC mass
distributions (defined in Section~\ref{subsec:categories}) for the $\tau$-embedded \Ztomumu\ data and simulated \Ztotautau\
events. For (a) only statistical uncertainties are shown; (b) also includes systematic
uncertainties associated with the embedding procedure as
discussed in Section~\ref{sec:systematics}.
\label{fig:embedding}}
\end{figure}
In a sample of selected \Ztomumu\ data 
events, the muon tracks and associated calorimeter cells are replaced by $\tau$ 
leptons from a simulated \Ztotautau\ decay with the same kinematics, where the $\tau$ 
polarisation and spin correlations are modelled with the TAUOLA program and the $\tau$-$\mu$ 
mass difference is taken into account as well. Thus, only the $\tau$ decays and the corresponding 
detector response are taken from the simulation, whereas the underlying event kinematics and 
all other properties---including pileup effects---are obtained from the data. 
These embedded data are used to model the shape of the relevant \Ztotautau\ background 
distributions as well as the efficiency for selecting \Ztotautau\ events from the 
preselected sample. The overall normalisation at the preselection level is obtained from 
simulation.

The procedure is extensively validated: example results for the $\Httlh$ channel are 
shown in Figure~\ref{fig:embedding}.
Systematic effects intrinsic to the method are studied by replacing the
muons selected in data by simulated muons instead of $\tau$
decays. Figure~\ref{fig:embedding}(a) shows a comparison of the \MET\ distributions from the 
selected \Ztomumu events in data before and after this muon embedding, demonstrating that the 
embedding procedure does not introduce any significant bias to the reconstruction of the 
event properties. Figure~\ref{fig:embedding}(b) compares the MMC mass 
distributions reconstructed from the $\tau$-embedded \Ztomumu\ data and simulated \Ztotautau\ 
events after the preselection described in Section~\ref{sec:preselection}; good agreement is 
found and similar studies for the other channels yield the same conclusions.

\boldmath
\subsection{\Httll}
\unboldmath
\label{subsec:bgll}

The $\Ztotautau$ background is modelled using the embedding procedure described above. 
The contribution from $Z/\gamma^{*}\rightarrow\ell^+\ell^-$ is
determined by scaling the yields in the Monte Carlo simulation using correction factors obtained by comparing data to simulation in low- and high-\MET\ control regions enriched in these backgrounds.
The correction factors are obtained separately for $\Ztoee$ and $\Ztomumu$ and for the different analysis categories and are on the order of 10\%.

The fake lepton
background consists of events that have a reconstructed lepton that did not originate from
the decay of a $\tau$ lepton or the leptonic decay of a 
$W$ or $Z$ boson. The
normalisation and shape of relevant distributions are obtained from data with a template
method using a control region in which the lepton isolation requirement is reversed. 
The chosen template shape is the \pt distribution of the sub-leading lepton. 
For this method to be applied, it is first verified that the template shapes of the fake lepton distribution in the control and signal regions agree within uncertainties. 
This is performed at intermediate steps of the event selection 
where the data sample is dominated by background events and where the number of expected signal events
is negligible.

After subtracting the simulated backgrounds, the template shape in a given distribution
is obtained from the control region, while the normalisation is obtained from a fit
of the distribution of the events in the signal region with the template shape. 
The uncertainty related to the
estimation of backgrounds with fake leptons is calculated from the uncertainty on the
subtraction of other processes from Monte Carlo simulation and from the difference in the \pt 
shape of the events  in the control region and signal regions.
Such systematic uncertainties lie in the range of 30--40\%.
\begin{table}[!t]
\begin{center}
\caption{Number of events after the \Httll\ selection for the four categories in data and predicted number of background events, for an integrated luminosity of 
$4.7\,\ifb$. Expectations for the Higgs boson signal ($m_{H} =120\gev$) are also given. Statistical and systematic uncertainties are quoted, in that order. \newline \
\label{tab:tautaullresults}}
\footnotesize
\begin{tabular}{lcccc}
\hline \hline
                                                  &  $ee$ + $\mu\mu$ + $e\mu$       & $ee$ + $\mu\mu$ + $e\mu$       & $ee$ + $\mu\mu$ + $e\mu$  &  $e\mu$  \\ 
                                                  &  $H+\!2$-jet VBF                & $H+\!2$-jet $VH$               & $H+\!1$-jet               & $H+\!0$-jet       \\
\hline 
$gg\to H$ signal                                    &  \phantom{0}0.26$\pm$0.06$\pm$0.10  & \phantom{0}0.8\phantom{0}$\pm$0.1\phantom{0}$\pm$0.2\phantom{0}     & \phantom{0}3.9\phantom{0}$\pm$0.2\phantom{0}$\pm$1.0\phantom{0}     &  23\phantom{.00}$\pm$1\phantom{.00}$\pm$3\phantom{.00}         \\
VBF $H$ signal                                      &  \phantom{0}1.08$\pm$0.03$\pm$0.11  & \phantom{0}0.10$\pm$0.01$\pm$0.01  & \phantom{0}1.15$\pm$0.03$\pm$0.01  &  \phantom{0}0.75$\pm$0.03$\pm$0.06 \\
$VH$ signal                                         &  \phantom{0}0.01$\pm$0.01$\pm$0.01  & \phantom{0}0.53$\pm$0.02$\pm$0.07  & \phantom{0}0.40$\pm$0.02$\pm$0.03  &  \phantom{0}0.52$\pm$0.02$\pm$0.04 \\
\hline
$Z/\gamma^{*} \to \tau^+\tau^-$                      & \phantom{0}24\phantom{.0}$\pm$\phantom{0}3\phantom{.0}$\pm$\phantom{0}2\phantom{.0}        & 107\phantom{.0}$\pm$12\phantom{.0}$\pm$\phantom{0}9\phantom{.0}     & (0.52$\pm$0.01$\pm$0.04)$\cdot 10^{3}$   &  (9.68$\pm$0.05$\pm$0.07)$\cdot 10^{3}$ \\
$Z/\gamma^{*} \to \ell^+\ell^-    $ ($\ell$=e,$\mu$) & \phantom{00}2\phantom{.0}$\pm$\phantom{0}1\phantom{.0}$\pm$\phantom{0}1\phantom{.0}         & \phantom{0}25\phantom{.0}$\pm$\phantom{0}4\phantom{.0}$\pm$\phantom{0}9\phantom{.0}       & \phantom{0}83\phantom{.0}$\pm$10\phantom{.0}$\pm$30\phantom{.0}         &  185\phantom{.0}$\pm$11\phantom{.0}$\pm$14\phantom{.0}      \\
$\ttbar$+single top                                 & \phantom{00}7\phantom{.0}$\pm$\phantom{0}1\phantom{.0}$\pm$\phantom{0}2\phantom{.0}         & \phantom{0}42\phantom{.0}$\pm$\phantom{0}2\phantom{.0}$\pm$\phantom{0}6\phantom{.0}       & \phantom{0}98\phantom{.0}$\pm$\phantom{0}3\phantom{.0}$\pm$12\phantom{.0}          &  169\phantom{.0}$\pm$\phantom{0}4\phantom{.0}$\pm$14\phantom{.0}       \\
$WW/WZ/ZZ$                                          & \phantom{00}0.9$\pm$\phantom{0}0.3$\pm$\phantom{0}0.3   & \phantom{00}6\phantom{.0}$\pm$\phantom{0}1\phantom{.0}$\pm$\phantom{0}1\phantom{.0}        & \phantom{0}21\phantom{.0}$\pm$\phantom{0}1\phantom{.0}$\pm$\phantom{0}3\phantom{.0}           &  221\phantom{.0}$\pm$\phantom{0}3\phantom{.0}$\pm$18\phantom{.0}       \\ 
Fake leptons                                        & \phantom{00}1.3$\pm$\phantom{0}0.8$\pm$\phantom{0}0.6   & \phantom{0}13\phantom{.0}$\pm$\phantom{0}2\phantom{.0}$\pm$\phantom{0}5\phantom{.0}       & \phantom{0}30\phantom{.0}$\pm$\phantom{0}4\phantom{.0}$\pm$12\phantom{.0}          &  (\phantom{0}1.2$\pm$0.5)$\cdot 10^{3}$    \\ 
\hline
Total background                                    & \phantom{0}35\phantom{.0}$\pm$\phantom{0}3\phantom{.0}$\pm$\phantom{0}4\phantom{.0}       & 193\phantom{.0}$\pm$\phantom{0}7\phantom{.0}$\pm$20\phantom{.0}       & (0.75$\pm$0.01$\pm$0.05)$\cdot 10^{3}$   &  (11.4$\pm$0.5)$\cdot 10^{3}$   \\ 
\hline
Observed data                                       &  27                         & 185                          & 702                        &  11420                  \\ 
\hline \hline

\end{tabular}
\end{center}
\end{table}
\begin{figure}[!t]
  \centering
  \subfigure[Jet multiplicity ($\pt>40$ GeV)]{
    \includegraphics[width=0.45\textwidth]{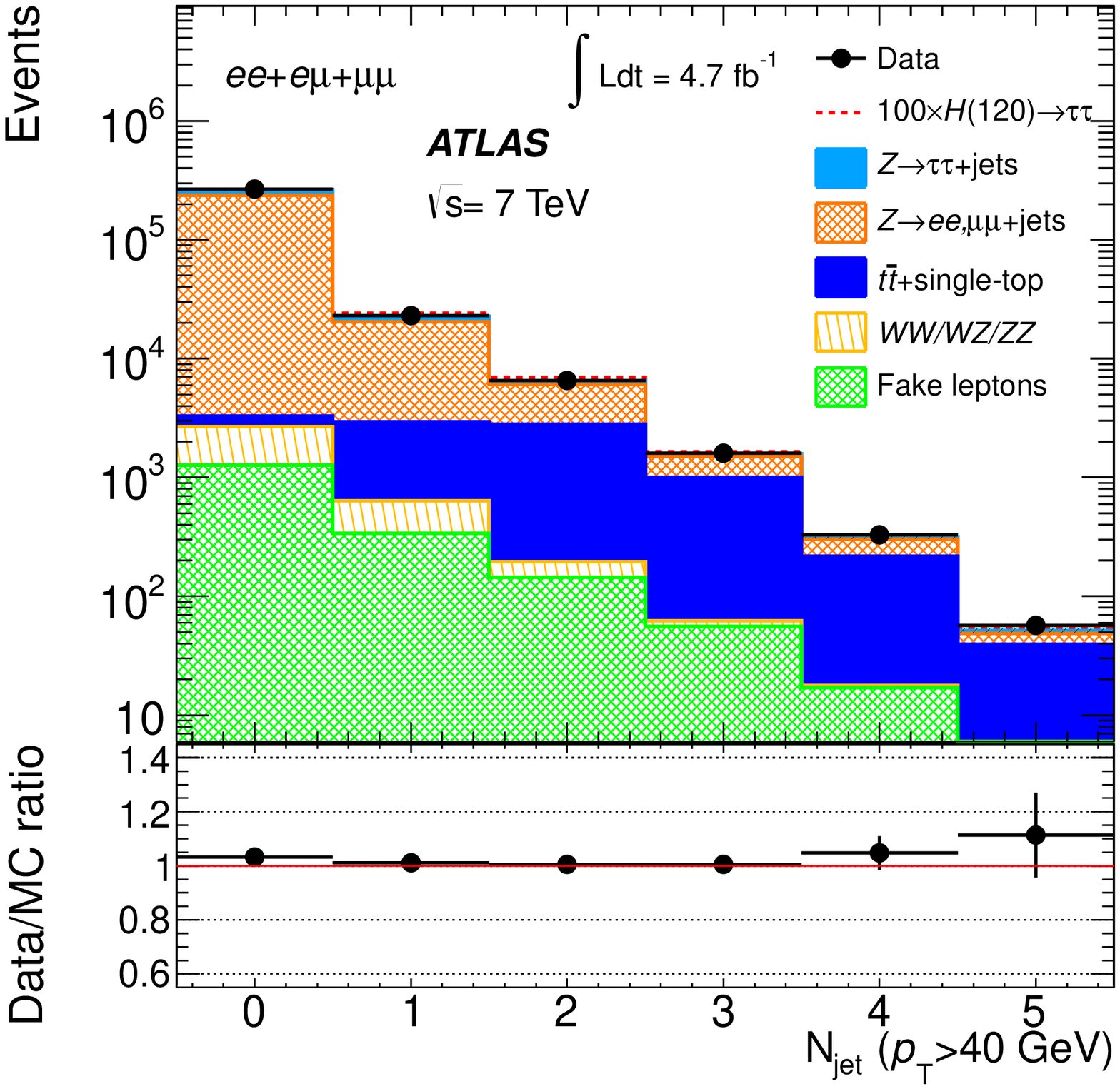}
  }
  \subfigure[\MET]{
    \includegraphics[width=0.45\textwidth]{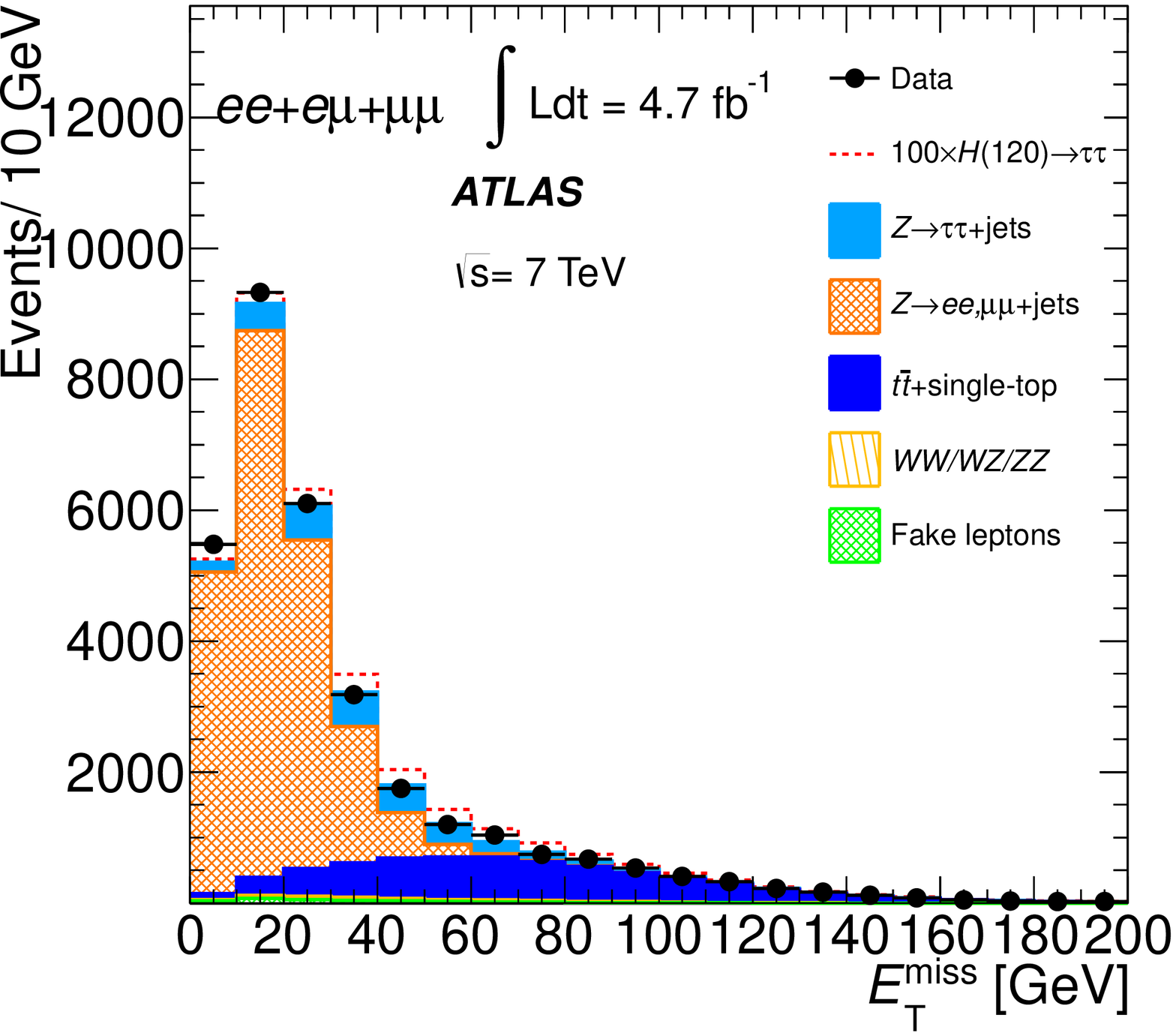}   
  }
  \subfigure[Invariant mass of the two leading jets]{
     \includegraphics[width=0.45\textwidth]{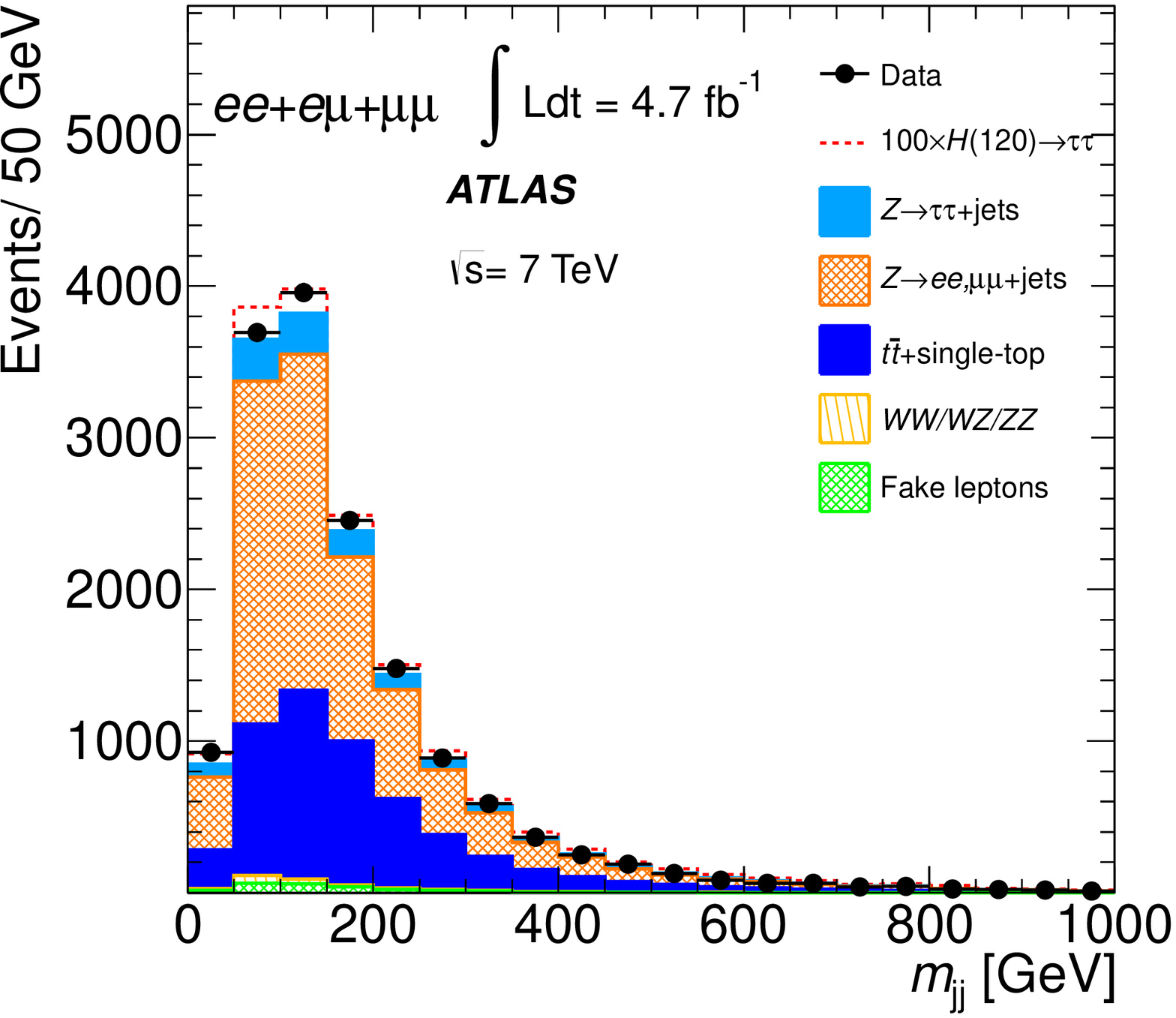}  
  }
 \subfigure[$\eta$ difference of the two leading jets]{
    \includegraphics[width=0.45\textwidth]{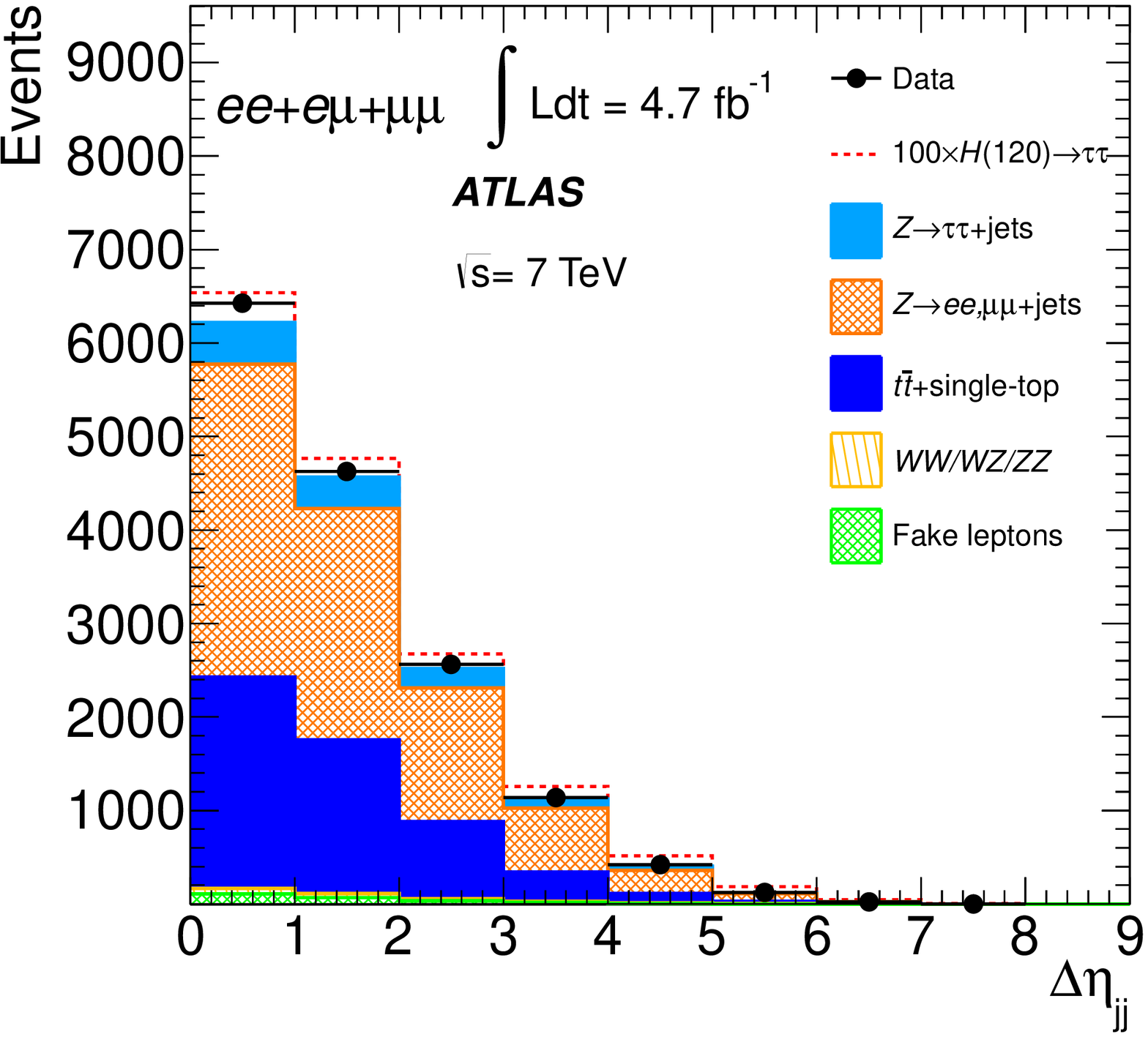}
  }
\caption[]{Distributions of (a) the jet multiplicity, (b) the \MET, 
(c) the invariant mass of the two leading jets and 
(d) the $\eta$ difference of the two leading jets  
in the \Httll\ channel for the selection criteria described in the text. 
Simulated 
samples are normalised to an integrated luminosity of $4.7\,\ifb$. For illustration only, the signal contributions have been scaled by factors given in the legends. 
$30\,\gev<m_{\ell\ell}<100\,\gev$ is required for the $ee$ and $\mu\mu$ channels and $30\,\gev<m_{\ell\ell}<75\,\gev$ for the $e\mu$ channel. For the \MET\ distribution the presence of a leading 
jet with \pt$>40~\gev$ is required; 
for the invariant mass and for the 
$\eta$ difference of the two leading jets, the presence of a sub-leading jet with \pt$>25~\gev$ is required in addition.
\label{fig:tautaulladd}}
\end{figure}
\begin{figure}[!t]
  \centering
  \subfigure[$H+\!0$-jet]{
    \includegraphics[width=0.45\textwidth]{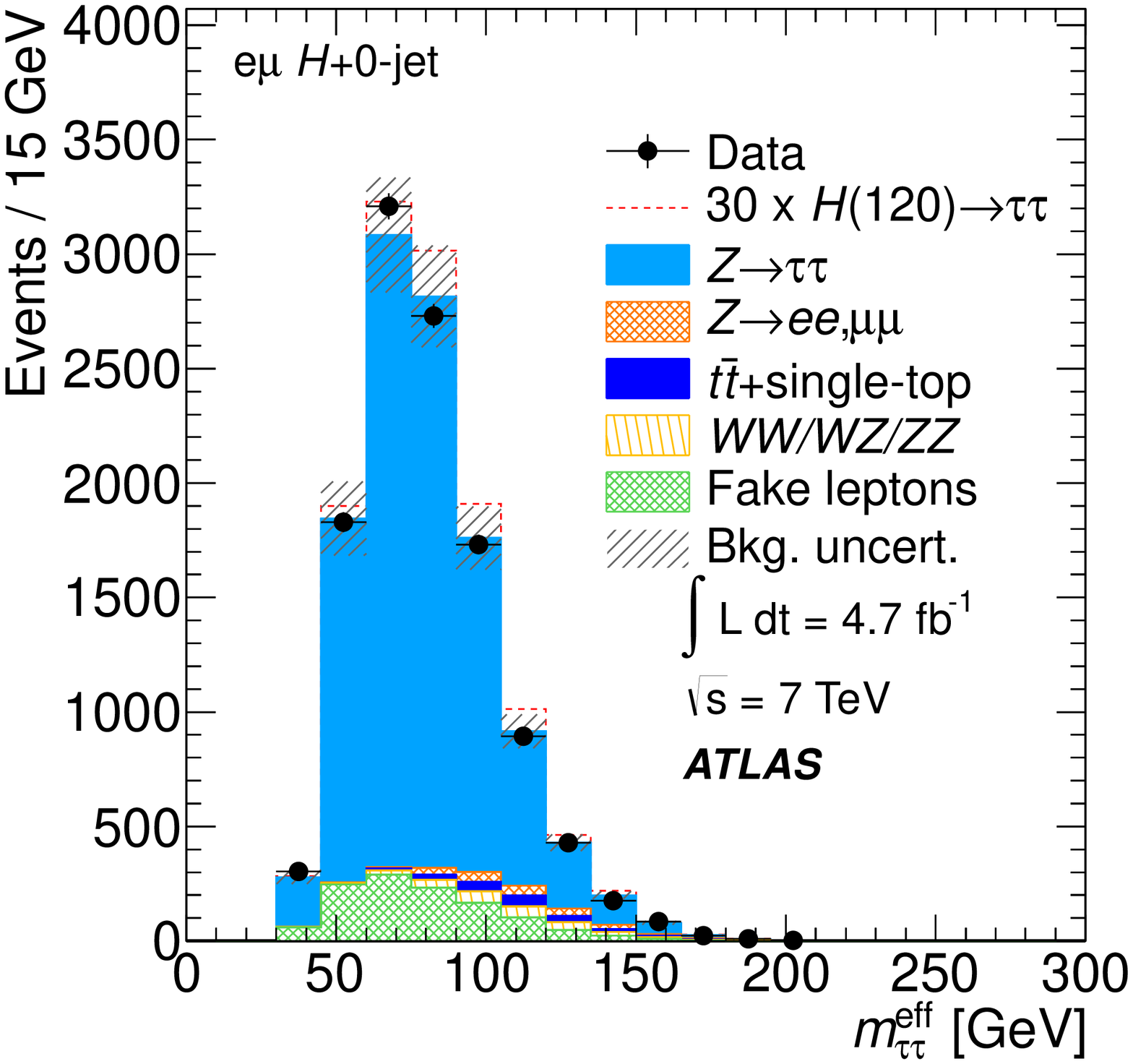}
  }
  \subfigure[$H+\!1$-jet]{
    \includegraphics[width=0.45\textwidth]{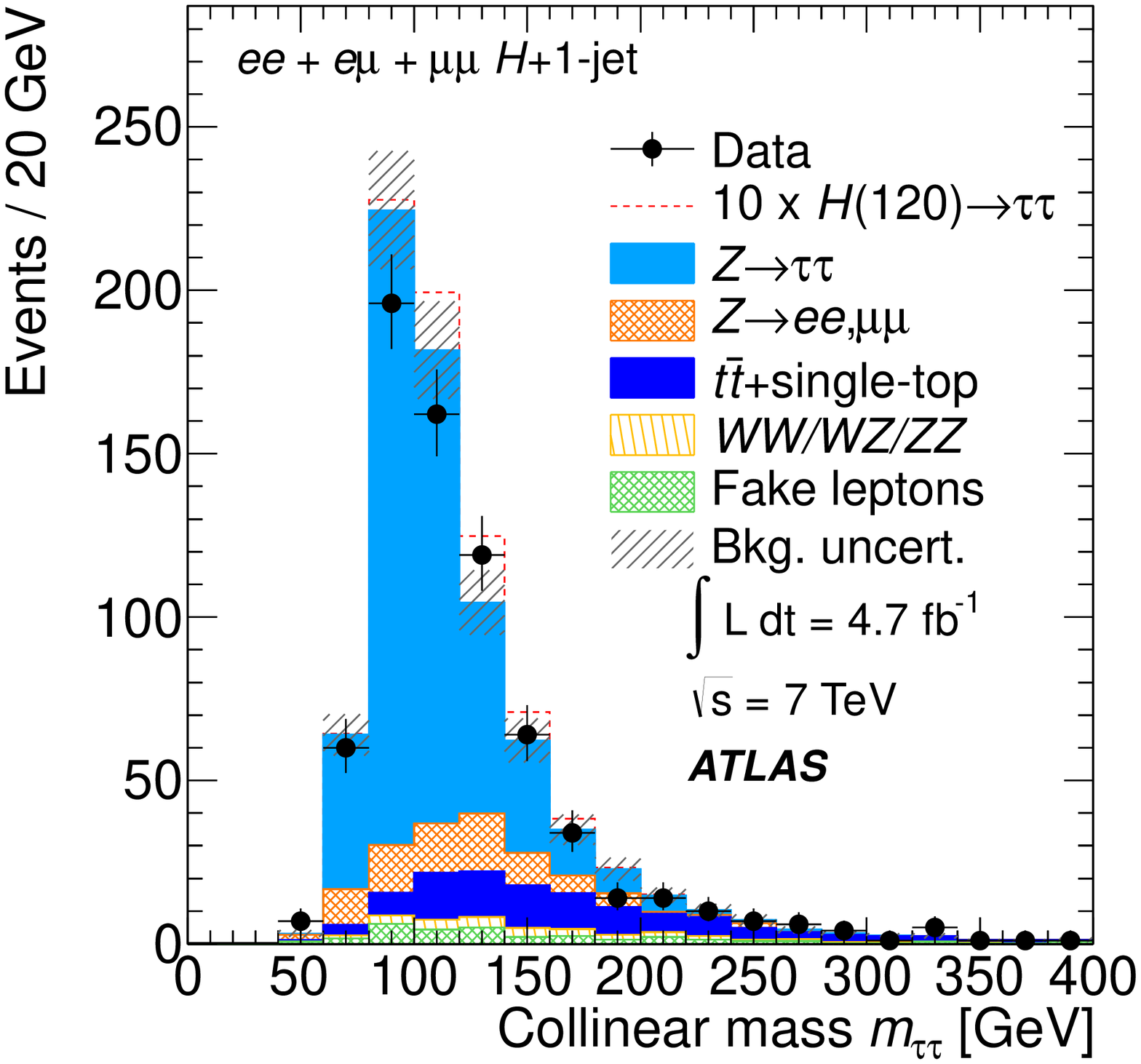}   
  }
 \subfigure[$H+\!2$-jet $VH$]{
    \includegraphics[width=0.45\textwidth]{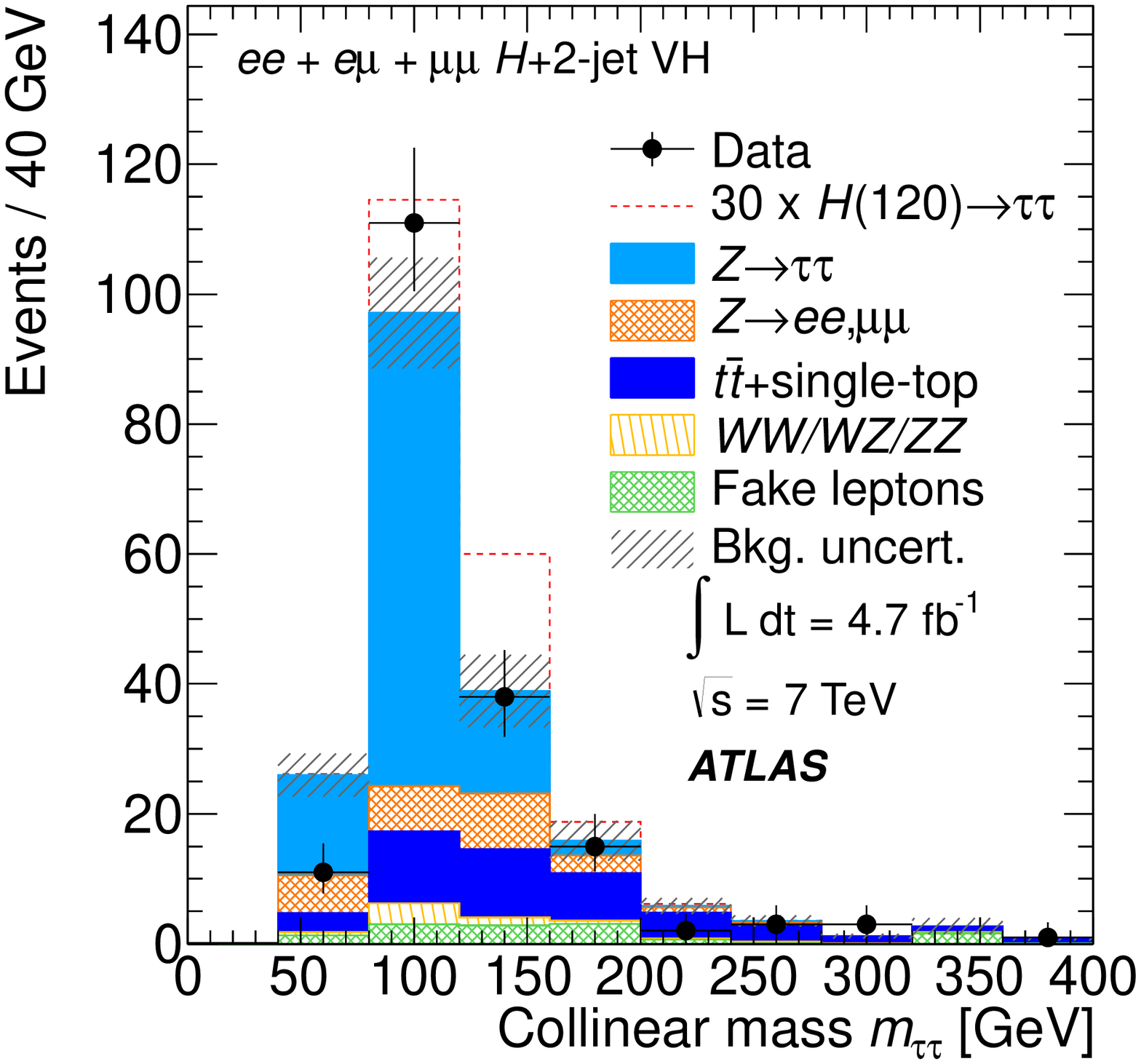}
  }
  \subfigure[$H+\!2$-jet VBF]{
     \includegraphics[width=0.45\textwidth]{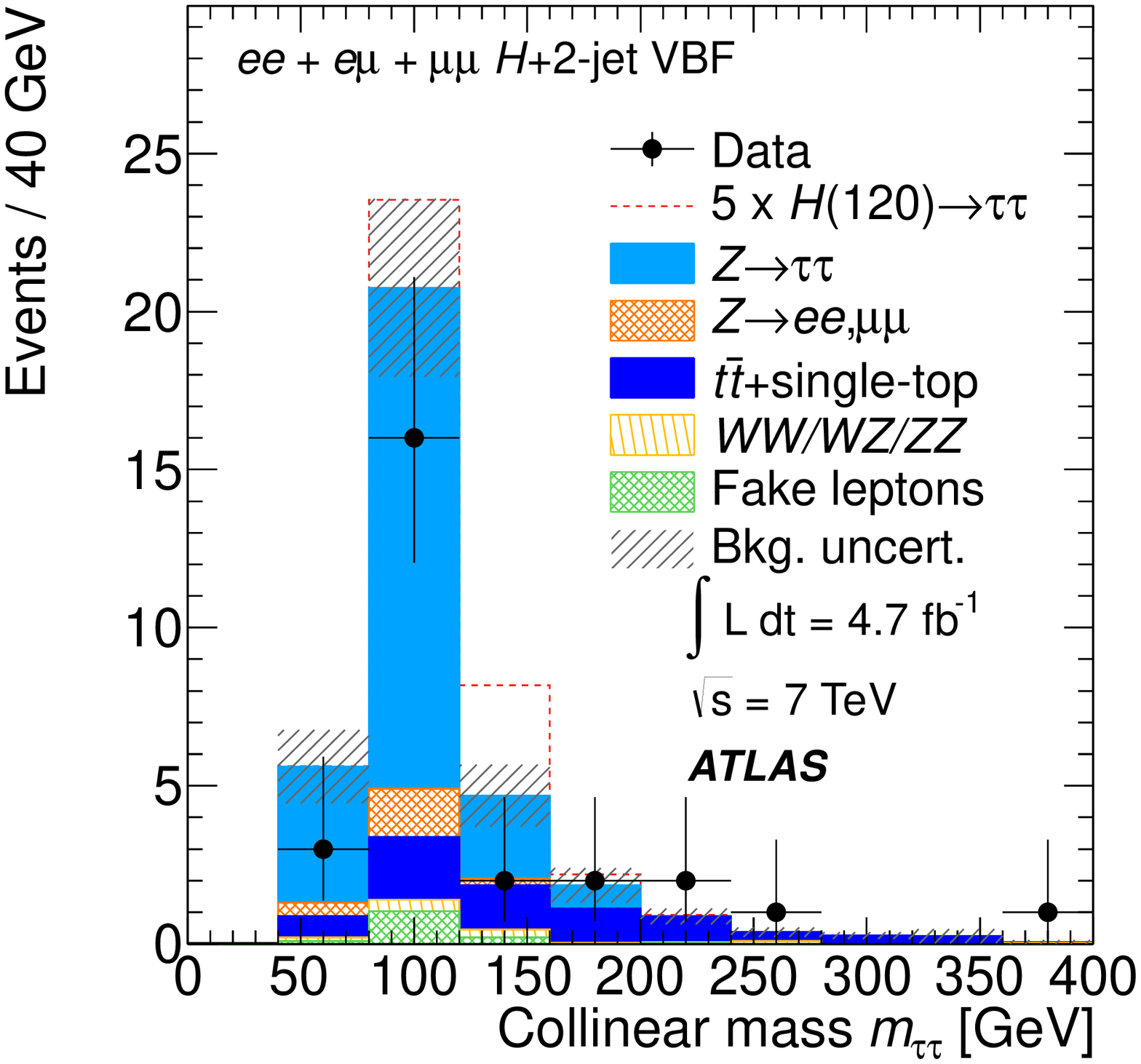}  
  }
\caption[]{Reconstructed $m_{\tau\tau}$ of the selected events in the \Httll\  channel for the four categories described in the text. Simulated samples are normalised to an 
integrated luminosity of $4.7\,\ifb$. Predictions from 
the Higgs boson signal ($m_{H} =120\gev$) and from backgrounds are given. In the case of the $H+\!0$-jet category \meff is used. 
For illustration only, the signal contributions have been scaled by factors given in the legends.
\label{fig:tautaullmh}}
\end{figure}

The contributions of the \ttbar, single top-quark and electroweak di-boson backgrounds are estimated from simulation. The Monte Carlo description of the top-quark backgrounds has been validated using data by 
selecting control regions enriched in top-quark background processes. The control regions are defined by inverting the $b$-jet selection for the $H+\!2$-jet VBF, $H+\!2$-jet $VH$, $H+\!1$-jet categories 
and by inverting the $H^{\textrm{lep}}_\mathrm{T}$ selection for the $H+\!0$-jet category.

Table~\ref{tab:tautaullresults} displays the number of events expected and observed in the four categories after all selection criteria including all systematic uncertainties described in 
Section~\ref{sec:systematics}. 
The estimated combined background contributions are found to give a good description of all 
quantities relevant to the analysis. As examples, the distributions of the jet multiplicity, \MET, the invariant mass of the two leading jets and  their pseudorapidity difference are shown in Figure~\ref{fig:tautaulladd}.
Figure~\ref{fig:tautaullmh} displays the invariant mass spectra of the selected events for the four categories.

\clearpage
\boldmath
\subsection{\Httlh}
\unboldmath
\label{subsec:bglh}

In order to estimate the background contributions to the selected 
\Httlh\ 
candidate events, a control sample in data is obtained by applying the signal selection 
described in Section~\ref{subsec:lhcat} but 
now requiring 
that the light lepton and the $\thad$ candidate have the same charge. This control sample 
is referred to as the same-sign (SS) sample here, in contrast to the 
opposite-sign (OS) signal sample.
The number of OS background events in the signal region (\nOS) can be expressed as
\begin{equation}
\label{eq:nOS_start}
  \nOS = \nSSall + \nOSSSwjets + \nOSSSzjets + \nOSSSother,
\end{equation}
where $\nSSall$ is the sum of all SS backgrounds in the signal region
and the remaining terms are the differences between the number of OS and SS
events for  \Wjets, \Ztotautau\ and other backgrounds, respectively. 
Due to their large production cross sections, multi-jet processes provide a significant background if quark/gluon jets 
are misidentified as hadronic $\tau$ decays.
The ratio of OS to SS events for the multi-jet background (\rqcd)
is expected to be  close to unity and therefore $\nOSSSqcd=0$ is assumed.
This assumption is validated with a control sample that is dominated by 
low-$\pt$ jets from multi-jet processes.
This sample is selected 
by replacing the requirement $\MET>20$~GeV with $\MET<15$~GeV and removing the isolation criteria of 
the electron or muon candidate. After subtraction of the other backgrounds using simulation, 
a value of $\rqcd = 1.10\pm 0.01 (\mathrm{stat.}) \pm 0.09 (\mathrm{syst.})$ is obtained. The 
observed 
deviation of \rqcd from unity is taken into account as a systematic 
uncertainty for the final result.
 
The  $\Ztotautau$ contribution is estimated from the  $\tau$-embedded $\Ztomumu$ sample 
in the data,
as described previously.
For the \Wjets background, a significant deviation of the ratio of OS and
SS events (\rwjets) from unity is
expected since \mbox{\Wjets} production is dominated by $gu$/$gd$-processes that often  give
rise to a jet originating from 
a quark, the charge of which is anti-correlated with the $W$ 
boson charge.
The predicted number of \Wjets background events is 
obtained from the simulation 
 after applying a normalisation correction factor  determined from $W$-dominated   
data control samples.\footnote{
These control samples are defined by replacing the $\mT < 30\GeV$ requirement in the nominal 
selection with $\mT>50$ GeV.  
}
The remaining contributions $\nOSSSother$ are taken from the simulation.

Table~\ref{tab:EventYield_SR} displays the number of events expected and observed in the 
seven categories after the full signal selection, including all systematic uncertainties as 
described in Section~\ref{sec:systematics}.
The estimated combined background contributions are found to give a good description of all 
quantities relevant to the analysis. As examples, the distributions of \MET, the transverse 
mass of the lepton-\MET system as well as the invariant mass of the two leading jets and 
their pseudorapidity difference are shown in Figure~\ref{fig:ControlPlots_lh}.
Figure~\ref{fig:MMCMass_AllCategory} shows the 
corresponding $\tau\tau$ invariant mass spectra, where the electron and muon categories have 
been combined for illustration purposes. The data are found to be consistent with the 
estimated combined background contributions in both normalisation and shape within the 
uncertainties. 
\begin{sidewaystable}[h!]
\begin{center}
\caption{
Predicted number of signal events (for $m_{H} =120\gev$) and predicted backgrounds obtained as described in the text, 
together with the observed number of events in data for the \Httlh\ categories. 
The total background  yield predicted by the alternative estimation method is given as well for comparison. The 
listed uncertainties are statistical and systematic, in that order. \newline \
\label{tab:EventYield_SR}
}
\small
\begin{tabular}{lc@{\ \ $\pm$\ }c@{\ $\pm$\ }cc@{\ \ $\pm$\ }c@{\  $\pm$\ }cc@{\ \ $\pm$\ }c@{\ $\pm$\ }cc@{\ $\pm$\ }c@{\ $\pm$\ }c}

\hline \hline
	             & \multicolumn{6}{c}{$H+\!0$-jet (low \MET)} & \multicolumn{6}{c}{$H+\!0$-jet (high \MET)}  \\
		     & \multicolumn{3}{c}{Electron} & \multicolumn{3}{c}{Muon} & \multicolumn{3}{c}{Electron} & \multicolumn{3}{c}{Muon} \\\hline 
$ggH$ signal        & \multicolumn{3}{c}{11\phantom{.0}\ $\pm$\ 1\phantom{0.0}\ $\pm$\ 2\phantom{0.0}}& \multicolumn{3}{c}{17\phantom{.0}\ $\pm$\ 1\phantom{0.0}\ $\pm$\ 4\phantom{0.0}}   & 
\multicolumn{3}{c}{7.1\phantom{0}\ $\pm$\ 0.8\phantom{0}\ $\pm$\ 1.5\phantom{0}}     & 
\multicolumn{3}{c}{9.8\phantom{0}\ $\pm$\ 0.9\phantom{0}\ $\pm$\ 2.1\phantom{0}}\\
VBF $H$ signal       & \multicolumn{3}{c}{0.08\ $\pm$\ 0.02\ $\pm$\ 0.12}  & \multicolumn{3}{c}{0.11\ $\pm$\ 0.03\ $\pm$\ 0.03}  & \multicolumn{3}{c}{0.09\ $\pm$\ 0.02\ $\pm$\ 0.02} & 
\multicolumn{3}{c}{0.14\ $\pm$\ 0.03\ $\pm$\ 0.03}\\
$VH$ signal         & \multicolumn{3}{c}{0.07\ $\pm$\ 0.02\ $\pm$\ 0.05} & \multicolumn{3}{c}{0.10\ $\pm$\ 0.03\ $\pm$\ 0.01}& \multicolumn{3}{c}{0.08\ $\pm$\ 0.0\ 2$\pm$\ 0.01}& 
\multicolumn{3}{c}{0.08\ $\pm$\ 0.02\ $\pm$\ 0.01}\\	    
\hline
    \nSSall\ & \multicolumn{3}{c}{$(3.3\phantom{0} \pm 0.2\phantom{0} \pm 0.7\phantom{0})\cdot10^3$} & \multicolumn{3}{c}{$(2.0\phantom{0} \pm  0.1\phantom{0} \pm  0.4\phantom{0})\cdot10^3$}& 
\multicolumn{3}{c}{$(0.69 \pm 0.06 \pm 0.14)\cdot10^3$} & 
\multicolumn{3}{c}{$(0.47 \pm 0.04 \pm 0.09)\cdot10^3$} 
\\      
    $\nOSSSwjets$ & \multicolumn{3}{c}{$(0.33 \pm 0.02 \pm 0.04)\cdot10^3$}  & \multicolumn{3}{c}{$(0.50 \pm 0.02 \pm 0.07)\cdot10^3$} & \multicolumn{3}{c}{$(0.15 \pm 0.01 \pm 0.02)\cdot10^3$}& 
\multicolumn{3}{c}{$(0.18 \pm 0.01 \pm 0.03)\cdot10^3$} \\
    $\nOSSSzjets$ & \multicolumn{3}{c}{$(3.70 \pm  0.06 \pm  0.61)\cdot10^3$} & \multicolumn{3}{c}{$(7.29 \pm 0.06 \pm 1.21)\cdot10^3$} & \multicolumn{3}{c}{$(1.49 \pm 0.04 \pm 0.23)\cdot10^3$} & 
\multicolumn{3}{c}{$(2.80 \pm 0.04 \pm 0.42)\cdot10^3$}\\  
    $\nOSSSother$ & \multicolumn{3}{c}{$(0.97 \pm 0.04 \pm 0.22)\cdot10^3$} & \multicolumn{3}{c}{$(0.59 \pm 0.04 \pm 0.14)\cdot10^3$} & \multicolumn{3}{c}{$(0.27 \pm 0.02 \pm 0.08)\cdot10^3$}      
& \multicolumn{3}{c}{$(0.14 \pm 0.02 \pm 0.04)\cdot10^3$}  
\\
\hline
Total background& \multicolumn{3}{c}{$(8.2\phantom{0} \pm 0.2\phantom{0} \pm  0.8\phantom{0})\cdot10^3$}  & \multicolumn{3}{c}{$(10.4\phantom{0}\pm 0.2\phantom{0} \pm 1.2\phantom{0})\cdot10^3$} & 
\multicolumn{3}{c}{$(2.59 \pm 0.07 \pm 0.26)\cdot10^3$} & \multicolumn{3}{c}{$(3.59 \pm 0.06 \pm 0.43)\cdot10^3$} \\
\hline
   Observed data  & \multicolumn{3}{c}{8363}& \multicolumn{3}{c}{10911}& \multicolumn{3}{c}{2545}& \multicolumn{3}{c}{3570} \\              
\hline \hline
Altern. estimate &  \multicolumn{3}{c}{$(8.7\phantom{0}\pm 0.1\phantom{0}\pm 0.8\phantom{0})\cdot10^3$}  &  \multicolumn{3}{c}{$(10.7\phantom{0} \pm 0.1\phantom{0} \pm 1.0\phantom{0})\cdot10^3$}  &  
\multicolumn{3}{c}{$(2.76 
\pm 0.05 \pm 0.33)\cdot10^3$}  &  
\multicolumn{3}{c}{$(3.75 \pm 0.05 \pm 0.47)\cdot10^3$}  \\
\hline\hline\\
& \multicolumn{12}{c}{\ } \\
\hline\hline
	             & \multicolumn{6}{c}{$H+\!1$-jet}  & \multicolumn{6}{c}{$H+\!2$-jet VBF} \\
		     & \multicolumn{3}{c}{Electron} & \multicolumn{3}{c}{Muon} & \multicolumn{6}{c}{Electron + Muon} \\\hline 
$ggH$ signal &\multicolumn{3}{c}{8.1\ $\pm$\ 0.7\ $\pm$\ 1.6} &\multicolumn{3}{c}{10.8\ $\pm$\ 0.8\ $\pm$\ 2.2}& \multicolumn{6}{c}{\phantom{0}0.9\phantom{0.} $\pm$ 0.2\phantom{0} $\pm$ 
0.3\phantom{0}}\\	
VBF$H$ signal&\multicolumn{3}{c}{1.6\ $\pm$\ 0.1\ $\pm$\ 0.1} &\multicolumn{3}{c}{\phantom{0}1.9\ $\pm$\ 0.1\ $\pm$\ 0.1}& \multicolumn{6}{c}{\phantom{0}2.2\phantom{0.} $\pm$ 0.1\phantom{0} $\pm$ 
0.2\phantom{0}}\\
$VH$ signal  &\multicolumn{3}{c}{1.1\ $\pm$\ 0.1\ $\pm$\ 0.1}&\multicolumn{3}{c}{\phantom{0}1.4\ $\pm$\ 0.1\ $\pm$\ 0.1}& \multicolumn{6}{c}{\phantom{0}0.02\phantom{.} $\pm$ 0.01 $\pm$ 
0.01} \\	 
\hline
\nSSall\ & \multicolumn{3}{c}{$(0.93\pm0.07\pm 0.19)\cdot10^3$}& \multicolumn{3}{c}{$(0.49\pm0.04\pm0.10)\cdot10^3$}& \multicolumn{6}{c}{\phantom{0}45\phantom{.} $\pm$ \phantom{0}7\phantom{.0} 
$\pm$ \phantom{0}9} 
\\		    
$\nOSSSwjets$ &\multicolumn{3}{c}{$(0.25\pm0.01\pm 0.03)\cdot10^3$}  &\multicolumn{3}{c}{$(0.26 \pm 0.01 \pm  0.03)\cdot10^3$}&\multicolumn{6}{c}{\phantom{00}5\phantom{.} $\pm$ 
\phantom{0}1\phantom{.0} 
$\pm$ \phantom{0}2} 
\\
$\nOSSSzjets$ &\multicolumn{3}{c}{$(1.23\pm0.03\pm0.17)\cdot10^3$}&\multicolumn{3}{c}{$(1.76\pm0.03\pm0.25)\cdot10^3$}&\multicolumn{6}{c}{\phantom{0}54\phantom{.} $\pm$ 
\phantom{0}6\phantom{.0} $\pm$ \phantom{0}8}\\ 
$\nOSSSother$ &\multicolumn{3}{c}{$(0.28 \pm 0.02 \pm 0.04)\cdot10^3$}&\multicolumn{3}{c}{$(0.24 \pm 0.01 \pm 0.03)\cdot10^3$}&\multicolumn{6}{c}{\phantom{0}20\phantom{.} $\pm$ 
\phantom{0}3\phantom{.0} $\pm$ 
\phantom{0}5}\\		
\hline
   Total background& \multicolumn{3}{c}{$(2.69 \pm 0.08 \pm 0.26)\cdot10^3$}& \multicolumn{3}{c}{$(2.75\pm 0.05 \pm 0.27)\cdot10^3$}& \multicolumn{6}{c}{124\phantom{.} $\pm$ 10\phantom{.0} $\pm$ 
13}\\		
\hline
   Observed data                    & \multicolumn{3}{c}{2610}& \multicolumn{3}{c}{2711}& \multicolumn{6}{c}{122} \\               
     \hline \hline
Altern. estimate &  \multicolumn{3}{c}{$(2.63 \pm 0.05 \pm 0.25)\cdot10^3$}  &  \multicolumn{3}{c}{$(2.72 \pm 0.04 \pm 0.28)\cdot10^3$}  & \multicolumn{6}{c}{$ 131\phantom{.} \pm 11\phantom{.0} 
\pm 23$}  
\\\hline\hline
 \end{tabular}
\end{center}
\end{sidewaystable}
\clearpage

\noindent

In order to validate these results, an alternative 
\begin{figure}[b!]
  \centering
  \subfigure[\MET]{
    \includegraphics[width=0.45\textwidth]{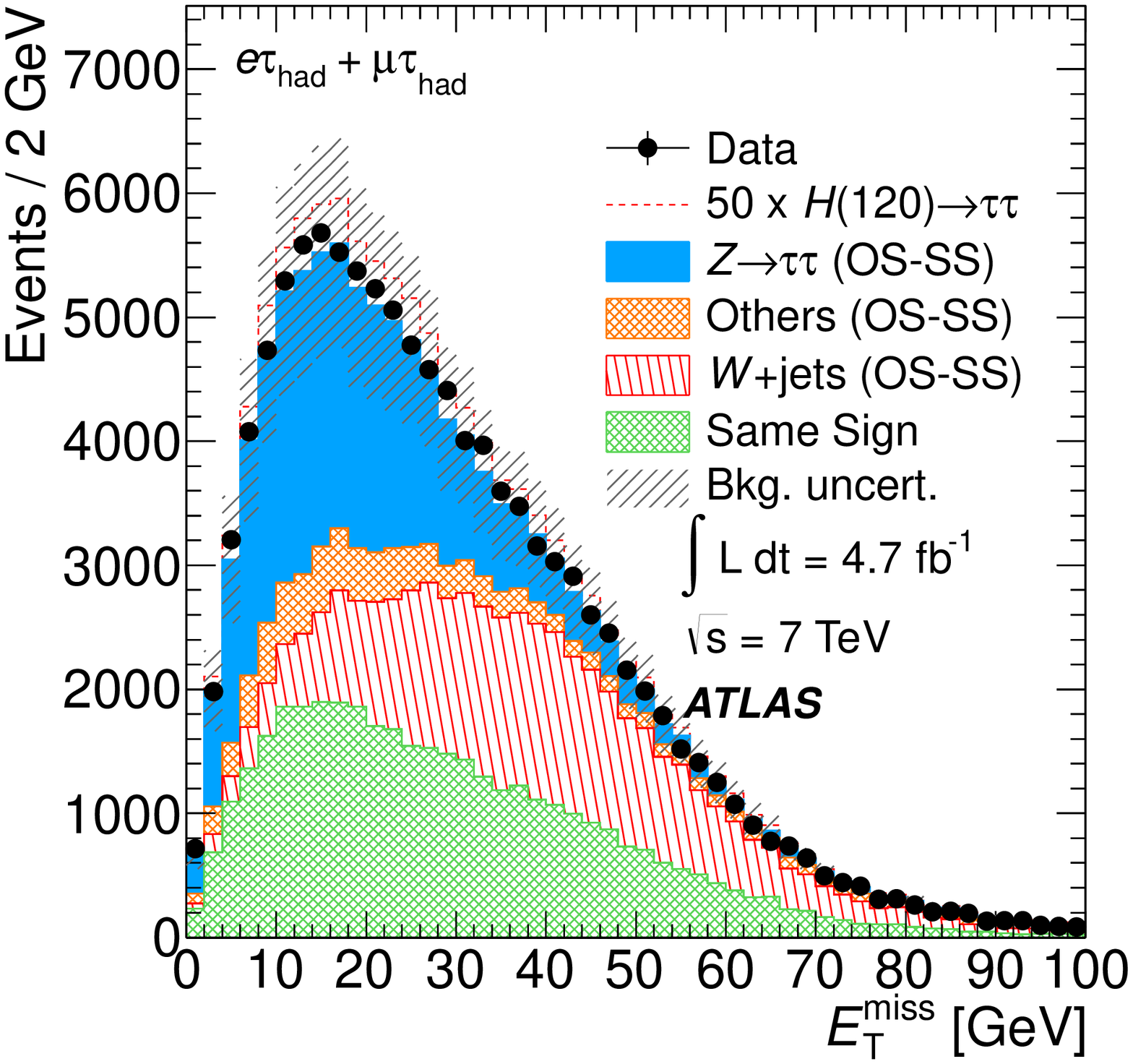}  
  }
  \subfigure[Transverse mass]{
    \includegraphics[width=0.45\textwidth]{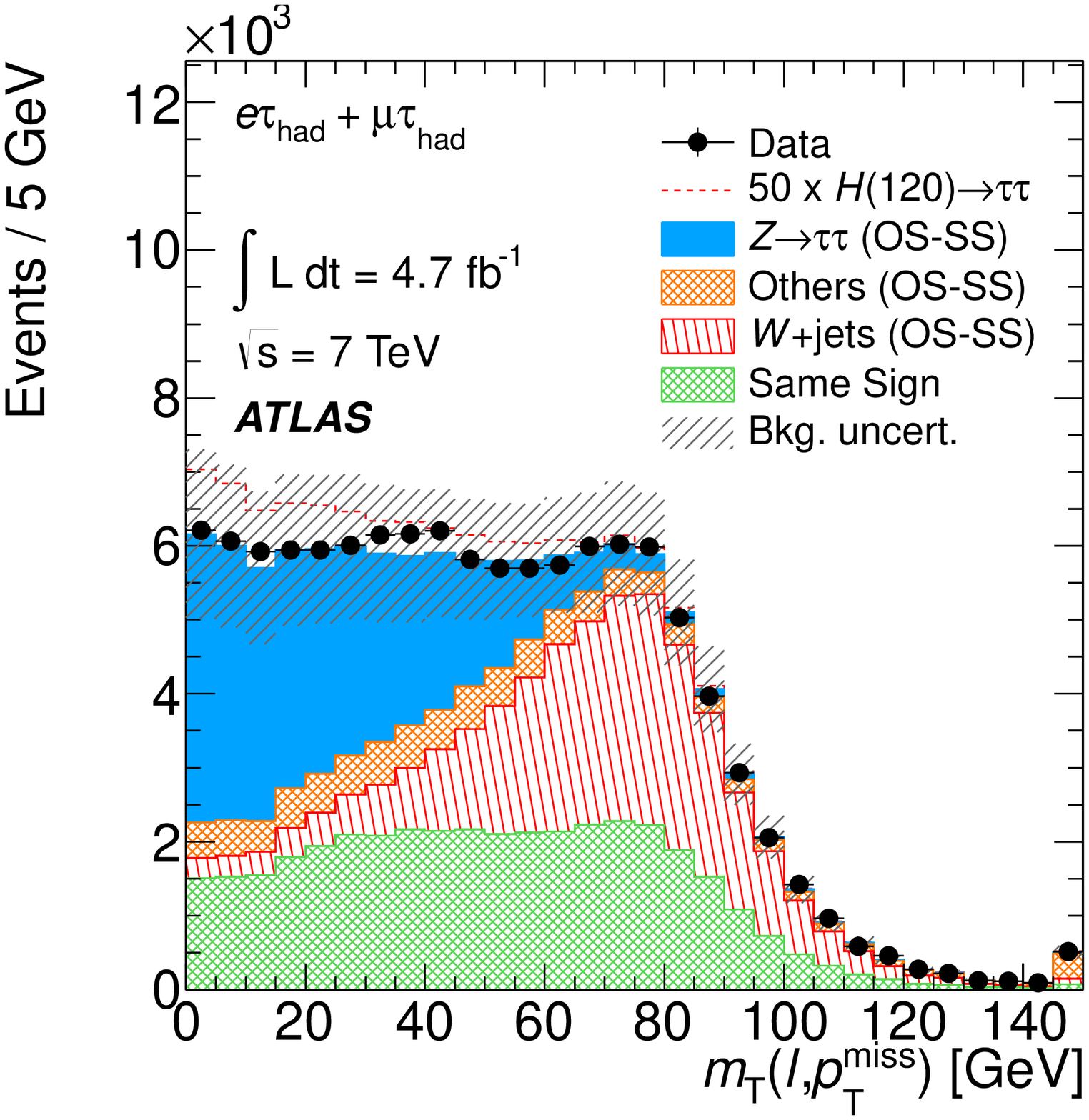}
  }
 \subfigure[Invariant mass of the two leading jets]{
    \includegraphics[width=0.45\textwidth]{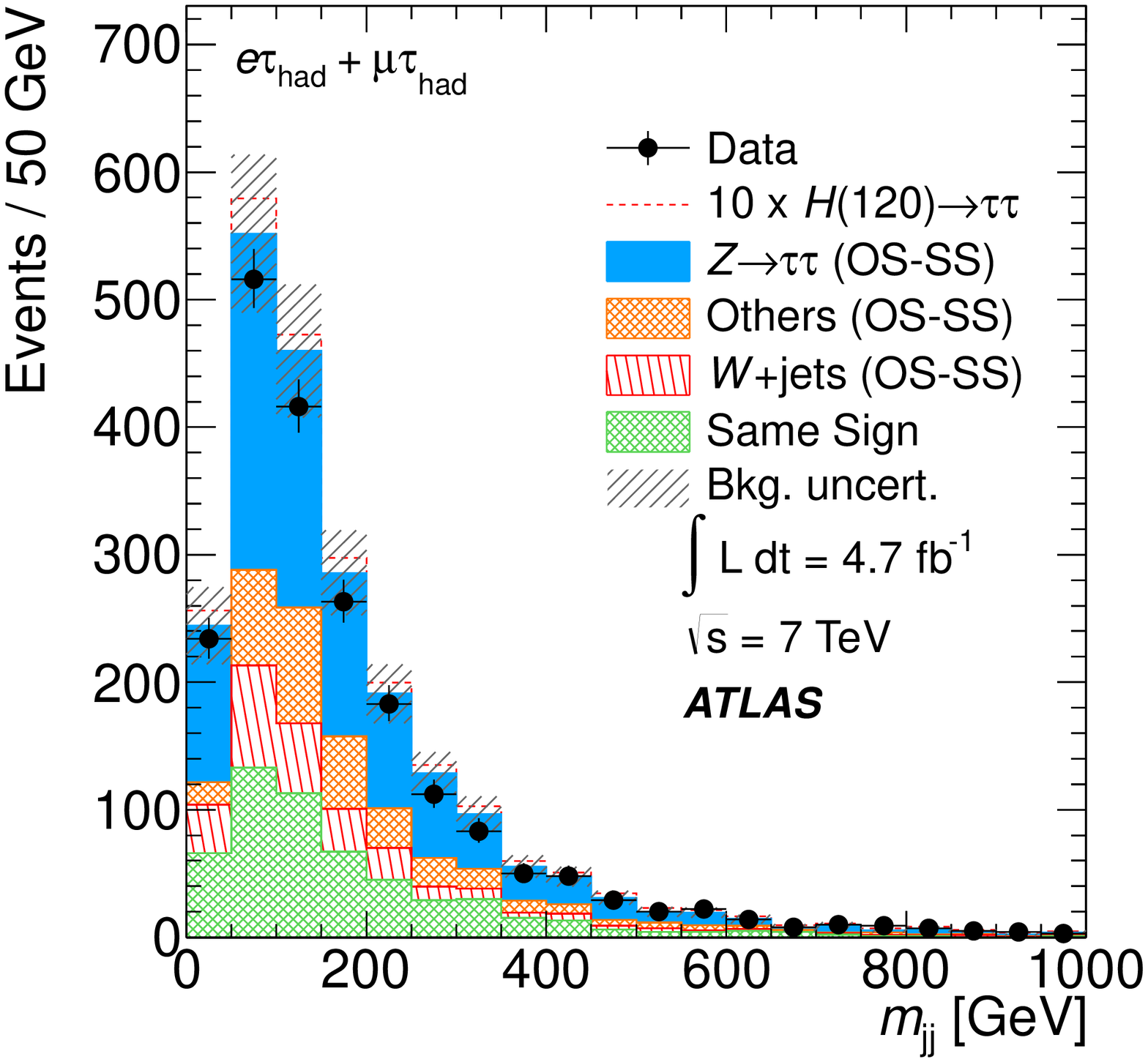}
  }
  \subfigure[$\eta$ difference of the two leading jets]{
    \includegraphics[width=0.45\textwidth]{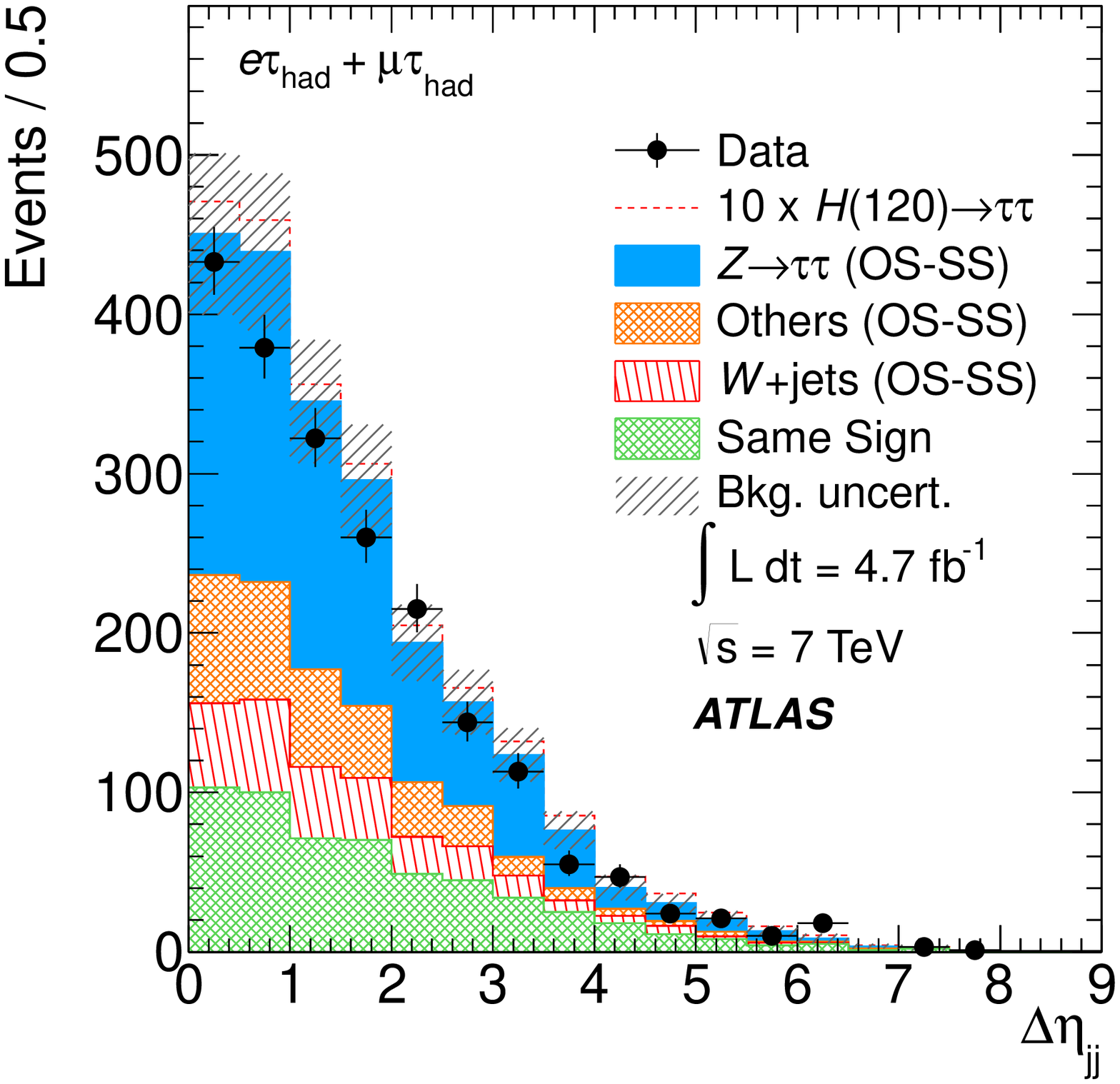}
  }
  \caption{
(a) Missing transverse momentum, (b) transverse mass of the lepton-\MET system as well as (c) invariant mass and (d) pseudorapidity difference of the two leading jets 
in the \Httlh\ channel. The preselection criteria and an $\MET>20\GeV$ requirement are applied except those for the displayed quantity; 
for (c) and (d) the presence of at least two jets with \pt$>25~\gev$ is required in addition.
Simulated samples are normalised to an integrated luminosity of $4.7\,\ifb$. The selected events in data are shown together with the predicted Higgs boson signal
($m_H=120\,\gev$) stacked above the background contributions (see text). For illustration only, the signal contributions have been scaled by factors given in the legends.
}
  \label{fig:ControlPlots_lh}
\end{figure}
background estimation is performed. This second method provides an estimate of the multi-jet 
background using control samples that 
are defined by separately inverting the light-lepton isolation and the $\thad$-$\ell$ charge correlation 
requirements, respectively. In addition to the signal region (A) this results in three 
background-dominated regions B (SS $\ell\thad$, $\ell$ isolated), C (OS $\ell\thad$, 
$\ell$ anti-isolated) and  D (SS $\ell\thad$, $\ell$ anti-isolated).
\begin{figure}[b!]
  \centering
  \subfigure[Combined $H+\!0$-jet (low \MET)]{
    \includegraphics[width=0.45\textwidth]{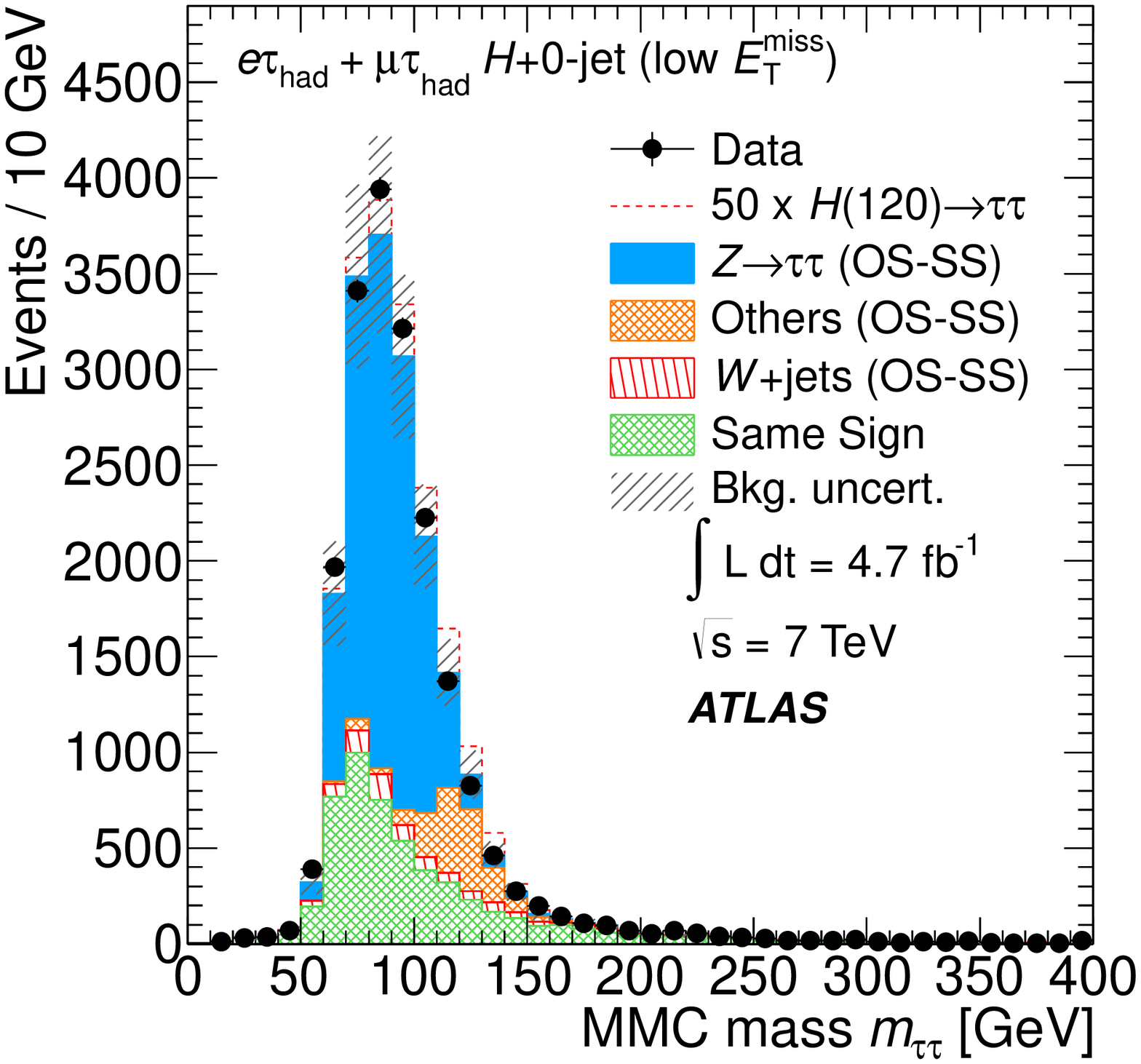}
  }
  \subfigure[Combined $H+\!0$-jet (high \MET)]{
    \includegraphics[width=0.45\textwidth]{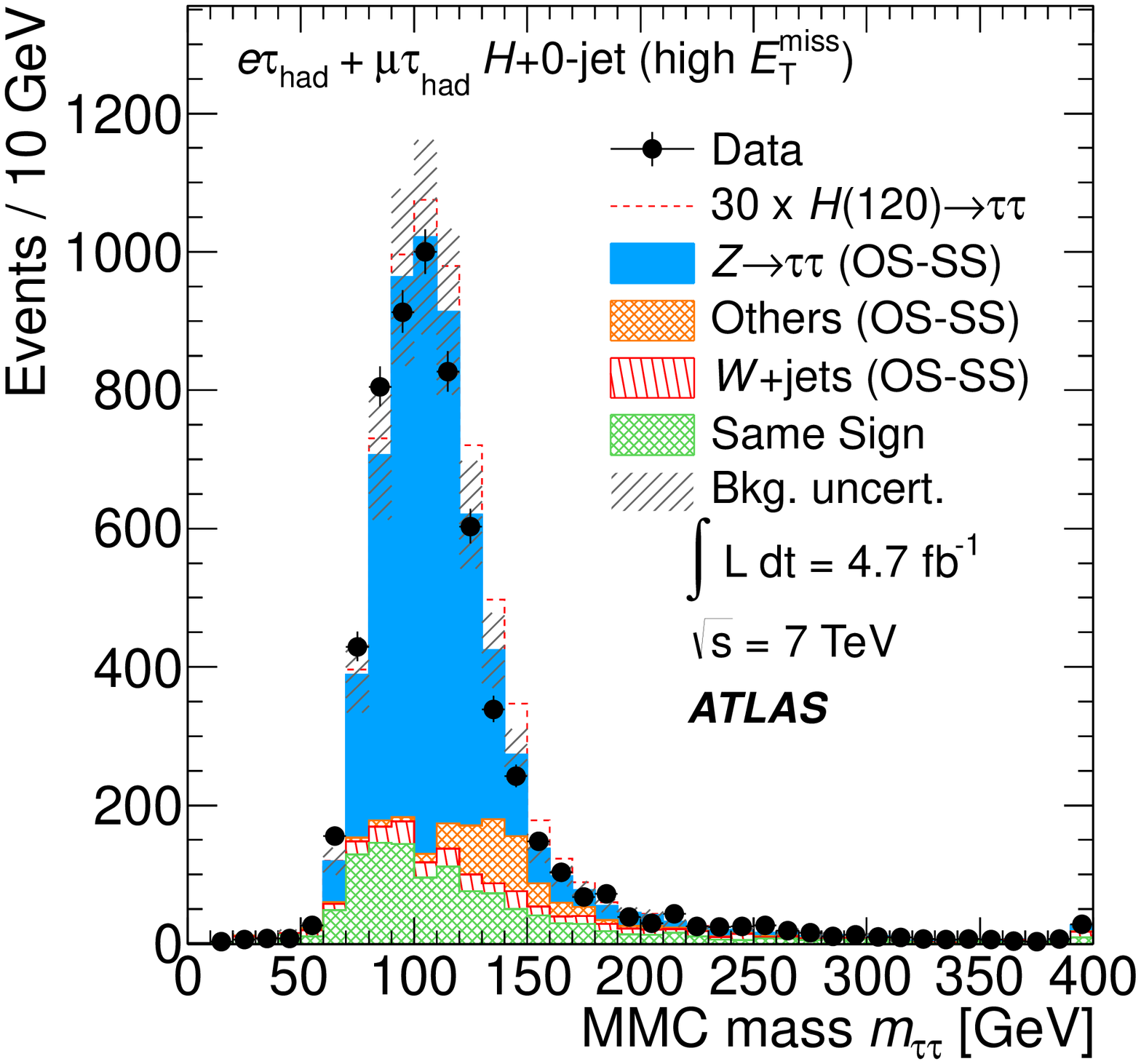}
  }
 \subfigure[Combined $H+\!1$-jet]{
    \includegraphics[width=0.45\textwidth]{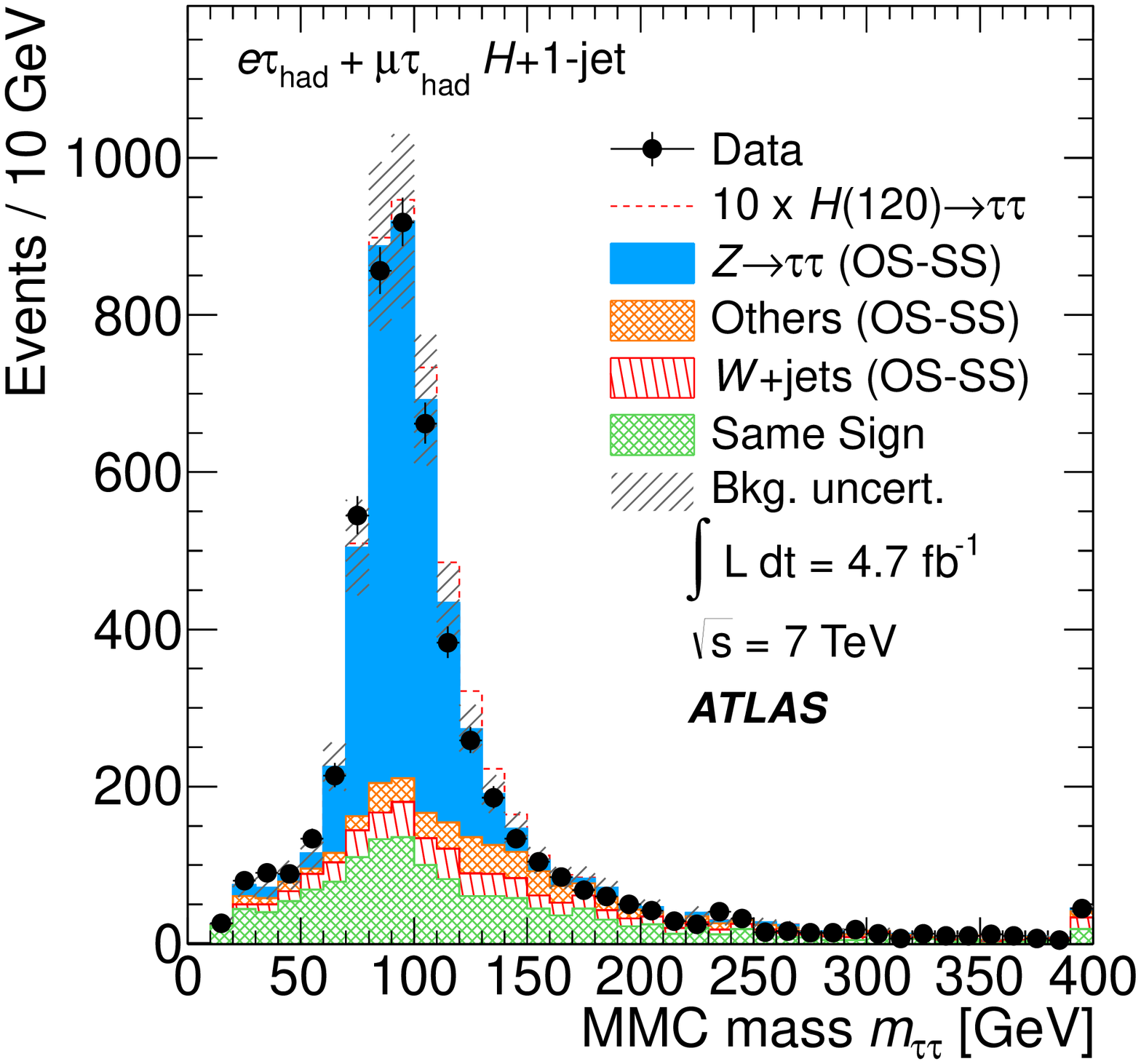}
  }
  \subfigure[$H+\!2$-jet VBF]{
    \includegraphics[width=0.45\textwidth]{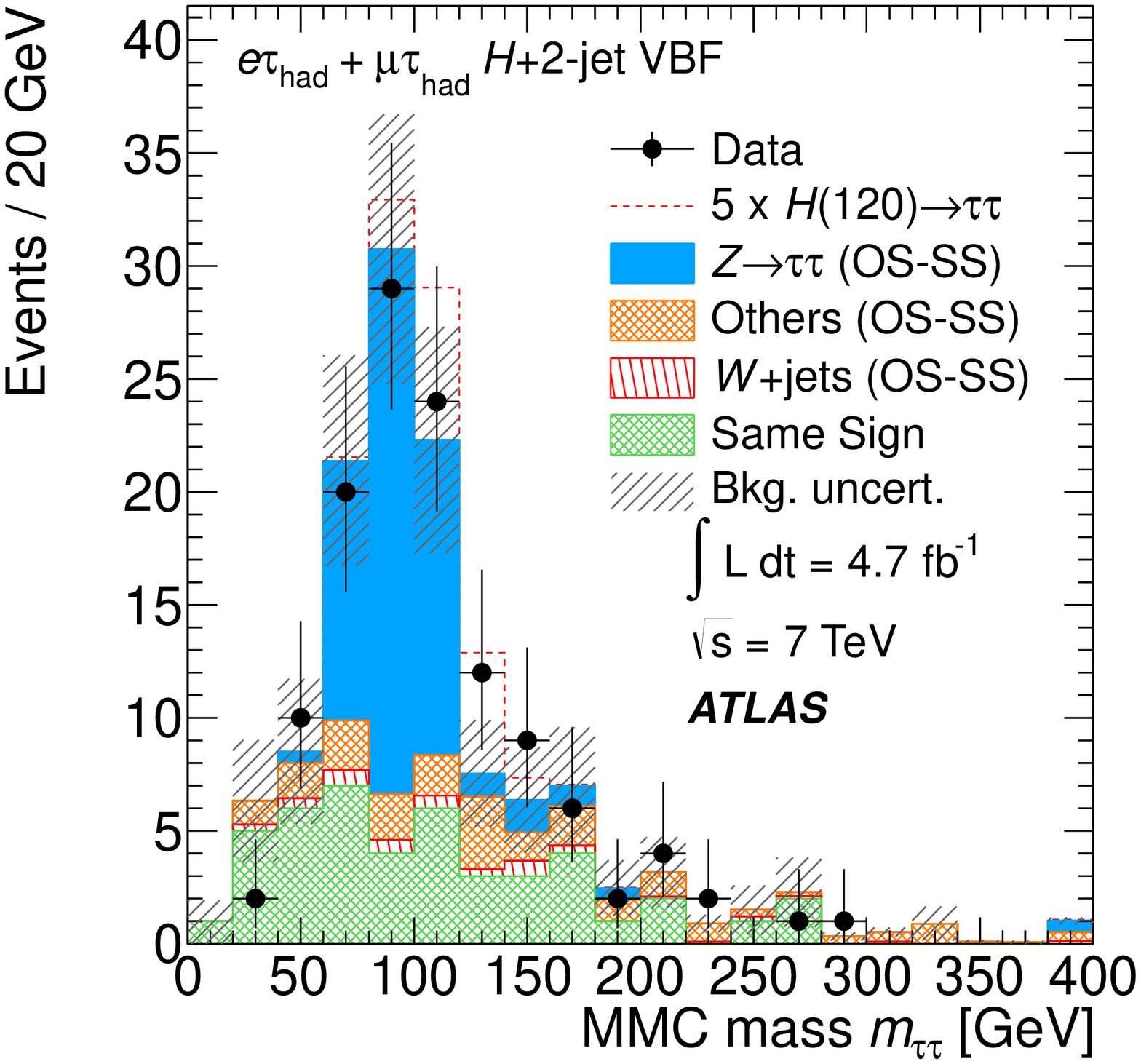}
  }
  \caption{
MMC mass  (defined in Section~\ref{subsec:categories}) distributions of the selected events in the \Httlh\ channel. The corresponding electron and muon categories 
for the $H+\!0$-jet and $H+\!1$-jet categories are shown combined here, while in the data analysis they are considered separately. 
The selected events in data are shown together with the predicted Higgs boson signal 
($m_H=120\,\gev$) stacked above the background 
contributions (see text). For illustration only, the signal contributions have been scaled by factors given in the legends.
}
  \label{fig:MMCMass_AllCategory}
\end{figure}
The shape of the MMC mass distribution for multi-jet contributions in the signal 
region A is taken from the control region~B and the normalisation is derived from the 
relation
$n_{\mathrm A}= n_\mathrm{C}/n_\mathrm{D} \times n_{\mathrm B}$. Here, $n_{\mathrm B}$, $n_{\mathrm C}$ and $n_{\mathrm D}$ denote the event yields in regions 
B, C and D, 
after subtracting the contribution from non-multi-jet backgrounds in all control regions. 
The $\Ztotautau$ contribution is estimated from the  $\tau$-embedded $\Ztomumu$ data and the estimate of the \Wjets background is obtained from the simulation 
after applying normalisation correction factors determined from $W$-dominated 
(high-\mT) OS and SS data control samples.
The remaining contributions are obtained from the simulation.
The resulting total background estimates are also included in Table~\ref{tab:EventYield_SR} 
and are found to be consistent with both the observed number of data events and the predictions of 
the default background estimate within the uncertainties.

Further studies of specific background contributions are performed by estimating 
the probability to misidentify jets as $\thad$ candidates 
in the signal region and using data control regions for the 
background from $t\bar{t}$ production processes. Results are consistent with the 
background estimates in Table~\ref{tab:EventYield_SR} in both cases.

\boldmath
\subsection{\Htthh}
\label{subsec:bghh}
\unboldmath

The dominant backgrounds in the \Htthh\ channel are \Ztotautau and multi-jet 
production. 
For both, the normalisation and shape of the mass distribution are estimated using data-driven methods.

The normalisation of the \Ztotautau background is obtained by using collision data events at an early stage of the event selection. 
This data-driven control sample is defined by requiring that events contain two $\thad$ candidates that pass the hadronic selections including the collinear approximation cuts (0 $<$ $x_{1,2}$ $<$ 
1 and $\Delta R(\tau\tau)<2.8$).
To avoid signal contamination in this control sample, a requirement that $m_{\tau\tau}$ $<$ 100 GeV is applied; this results in a
SM Higgs signal contamination of less than 0.2\%. 
The \Ztotautau contribution is obtained by fitting the track multiplicity of the two $\tau$ candidates simultaneously.
The tracks associated to the $\thad$ candidates are counted in the cone defined by $\Delta R < 0.6$, as motivated by the  
momentum correlation between tracks in $\thad$ candidates~\cite{TauID}.
A two-dimensional distribution of the track multiplicities for these two $\tau$ candidates is formed, and a track multiplicity fit is performed.
The multi-jet template is modelled from
the same-sign (SS) candidates in the data while the $\Ztotautau$ contribution is  modelled by the simulation. 
The less significant backgrounds are estimated from the simulation and subtracted before the  fit is performed.
The result of the fit is  used to normalise the $\tau$-embedded $\Ztomumu$ sample described above, and then this  sample 
is used to model the acceptance of the later cuts and the mass shape in the signal region.

The multi-jet contribution is 
estimated by the same two-dimensional track multiplicity fitting technique, where the fit is now performed
in the signal region. The contribution from di-$\tau$ events is allowed to float in the fit, where the shape comes from the simulation.
It is assumed that the shape of the two-dimensional track multiplicity in the $\Ztotautau$ and Higgs boson signal processes are the same.
\begin{figure}[b!]
  \centering
  \subfigure[$\Delta$R of the two $\tau$ candidates]{
    \includegraphics[width=0.45\textwidth]{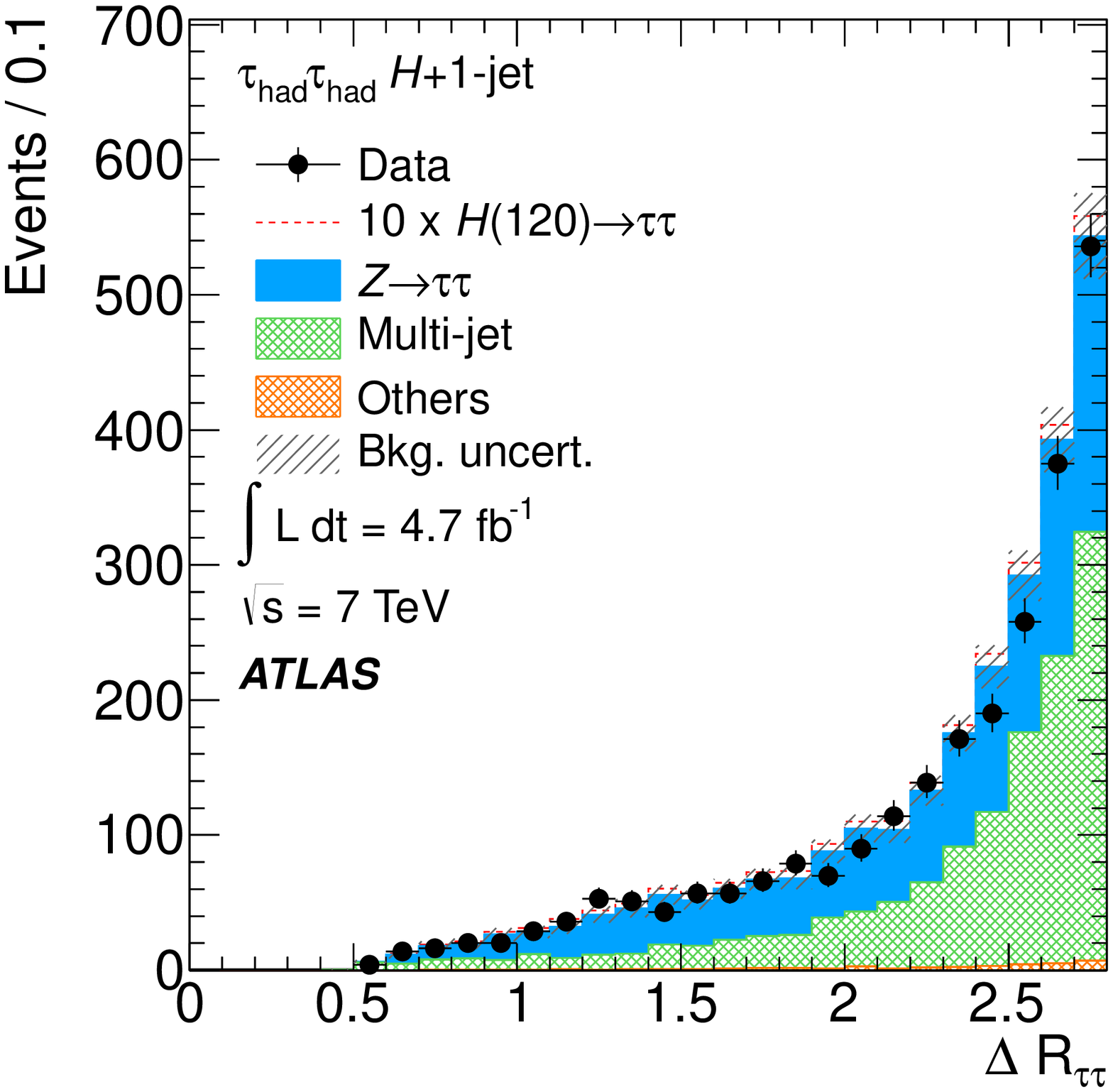}
  }
  \subfigure[\MET]{
    \includegraphics[width=0.45\textwidth]{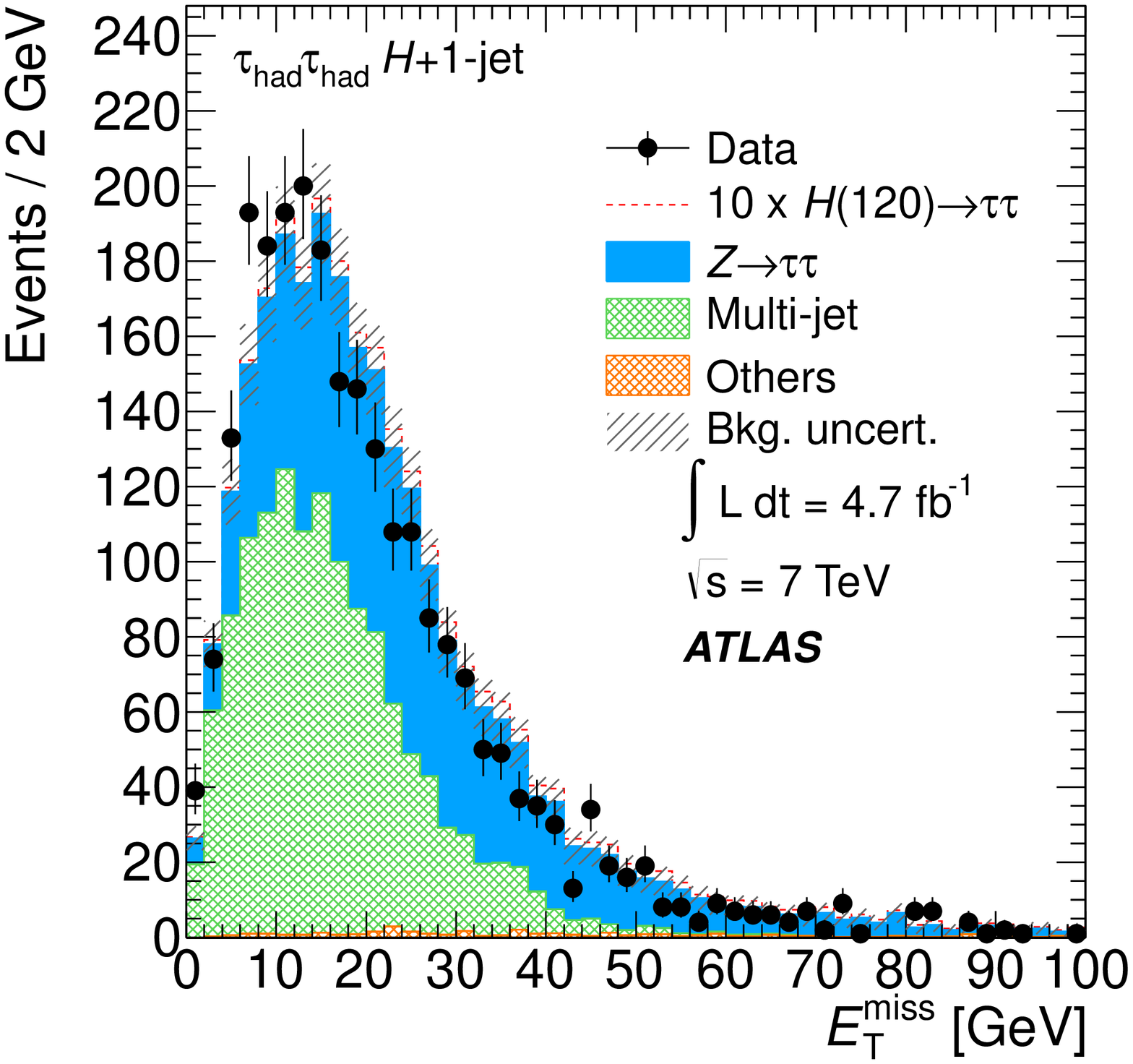}
  }
 \subfigure[Collinear mass]{
    \includegraphics[width=0.45\textwidth]{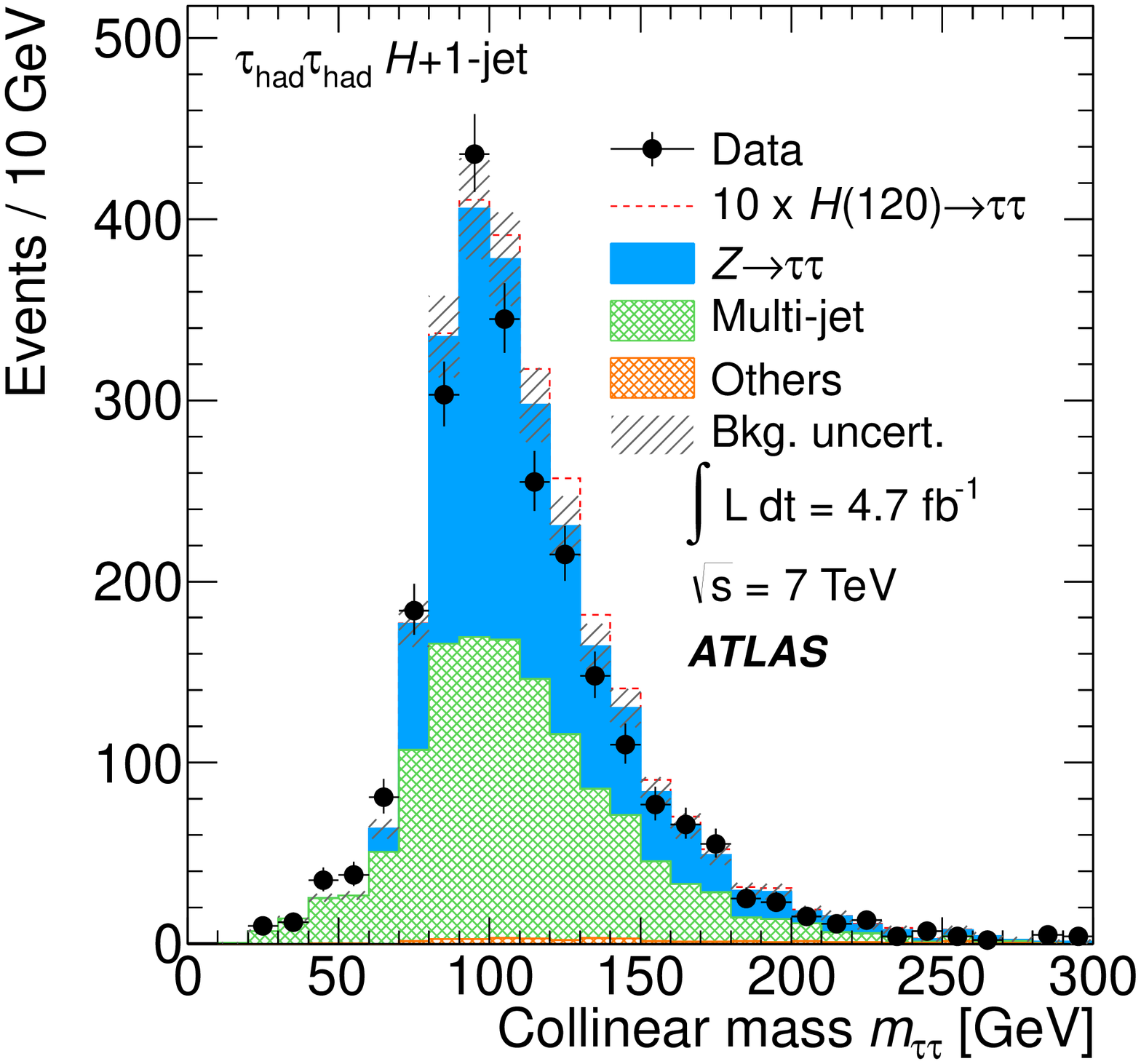} 
  }
  \subfigure[Invariant mass $m(H,\mathrm{jet1})$]{
    \includegraphics[width=0.45\textwidth]{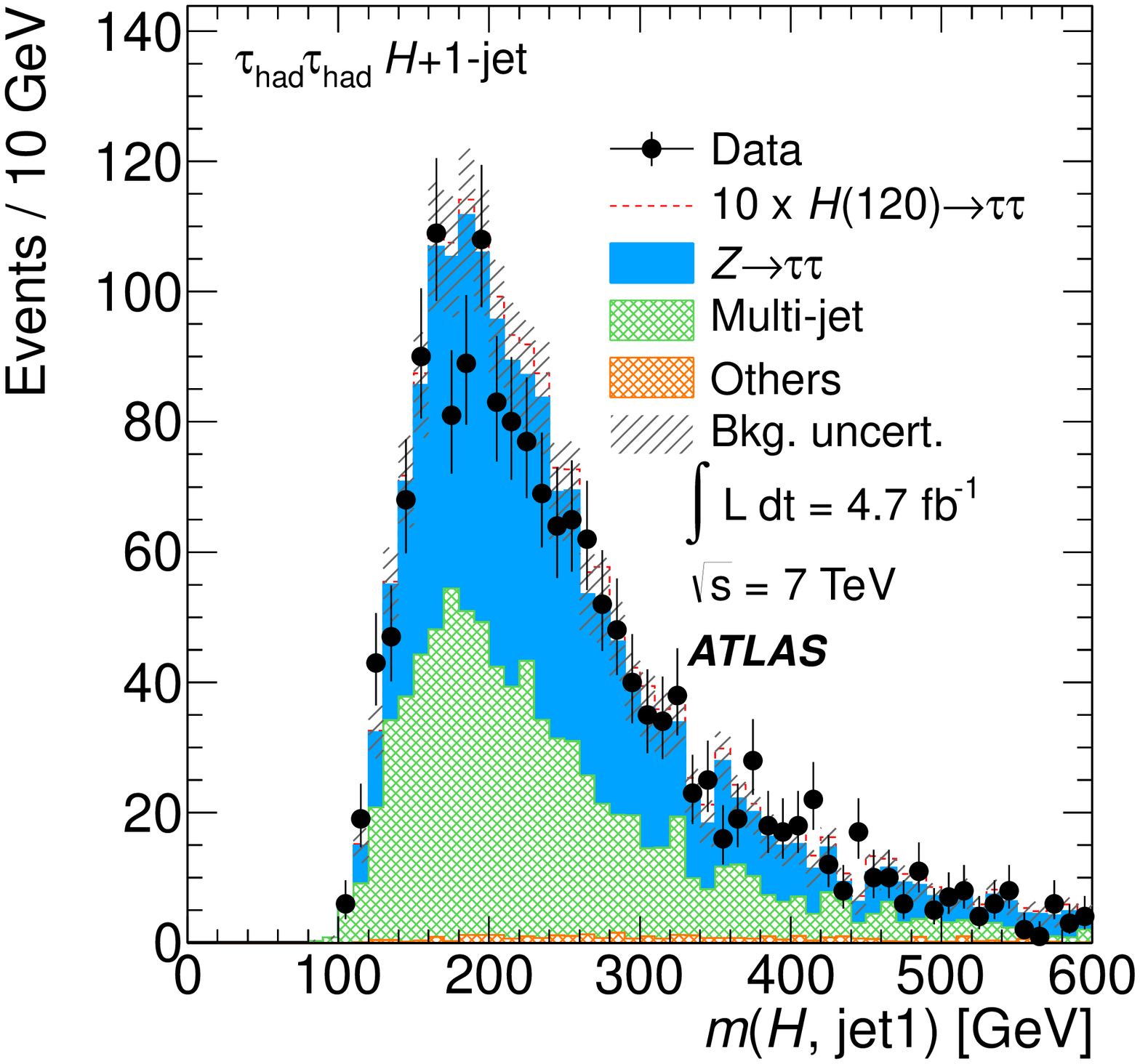}
  }
  \caption{
(a) $\Delta R$ of the two $\tau$ candidates, (b) missing transverse momentum, 
(c) collinear mass and (d) invariant mass of the reconstructed Higgs boson and the leading jet
in the \Htthh\ control region.
Expectations from the Higgs boson signal ($m_H=120\,\gev$)  and from backgrounds are given.
The error band indicates the total background uncertainty.
For illustration only, the signal contributions, which are expected to be small in this control region, have been scaled by factors given in the legends.
}
\label{fig:ControlPlots_hh}
\end{figure}

The systematic uncertainties considered for the background prediction arise from the jet template statistics, alternative 
multi-jet track multiplicity templates, the presence of non-di-$\tau$ background, signal contamination, charge misidentification probability and variations in the pileup conditions. 

Instead of using the SS events for the multi-jet track template, the alternative 
multi-jet track multiplicity template is built from events with one additional 
electron or muon which enhances the contribution of $W$+jets events, where two jets are 
misidentified as the hadronic taus. 
This provides a different mix of quark- and gluon-initiated jets from the inclusive multi-jet sample, addressing a possible flavour dependence of the track multiplicity distribution.

The overall systematic uncertainties are 11.6\% and 22\% for \Ztotautau and multi-jet backgrounds, respectively.

The mass shape from the multi-jet events is modelled by dropping the opposite-sign and $\thad$ track multiplicity requirements.
In order to obtain a data sample that is enriched with the multi-jet background, events are accepted if they are same-sign events or if the sum of the charges of the products of a single $\tau$ decay is 0 or $\pm2$.
This mass shape model is tested against several other hypothesis obtained, e.g., from pure SS samples or events selected with looser $\tau$ identification criteria.

Figure~\ref{fig:ControlPlots_hh} shows the kinematic distributions of (a) the $\Delta R$ of the two $\tau$ candidates, (b) the missing transverse momentum,
(c) the collinear mass and (d) the invariant mass of the reconstructed Higgs boson and the leading 
jet in the \Htthh\ control region.

Table \ref{table:nominalCutflow} presents the event yields after the full event selection, where the yields are normalised 
to $4.7\,\ifb$. 
The collinear mass distribution after full selection is presented in Figure~\ref{colmassSR}. 

\begin{table}[htdp]
\begin{center}
\caption{Number of events after the \Htthh\ selection in data 
and predicted number of background events, 
for an integrated luminosity of $4.7\,\ifb$. Predictions for  the Higgs boson signal ($m_{H} =120\gev$) are also given. 
Statistical errors and systematic uncertainties are quoted, in that order.\newline \
\label{table:nominalCutflow}}
\begin{tabular}{lc}
\hline \hline
                                 &  \Htthh \\ 
                                 &  $H+\!1$-jet  \\
\hline 
$gg\to H$ signal                 & 3.1\phantom{00}  $\pm$ 0.2\phantom{0}  $\pm$ 0.6\phantom{0}  \\
VBF $H$ signal                   & 1.51\phantom{0} $\pm$ 0.05 $\pm$ 0.18 \\
$VH$ signal                      & 0.61\phantom{0} $\pm$ 0.04 $\pm$ 0.06 \\
\hline
$Z/\gamma^{*} \to \tau^+\tau^-$   & 287\phantom{.0}   $\pm$ 23\phantom{.0}   $\pm$ 34\phantom{.0} \\
$W$+jets / $Z$+jets              & \phantom{00}6.1 $\pm$  \phantom{0}1.3 $\pm$  \phantom{0}1.4 \\
$\ttbar$+single top              &  \phantom{00}1.9 $\pm$  \phantom{0}0.3 $\pm$  \phantom{0}0.4 \\
$WW/WZ/ZZ$                       &  \phantom{00}2.1 $\pm$  \phantom{0}0.4 $\pm$  \phantom{0}0.4 \\ 
Multi-jet                        &  \phantom{0}54\phantom{.0}   $\pm$ 21\phantom{.0}   $\pm$ 12\phantom{.0} \\ 
\hline
Total background                 & 348\phantom{.0}   $\pm$ 31\phantom{.0}   $\pm$ 36\phantom{.0} \\ 
\hline
Observed data                    & 317                     \\ 
\hline \hline
\end{tabular}
\end{center}
\end{table}

\begin{figure}[htbp]
\begin{center}
\includegraphics[width=8cm]{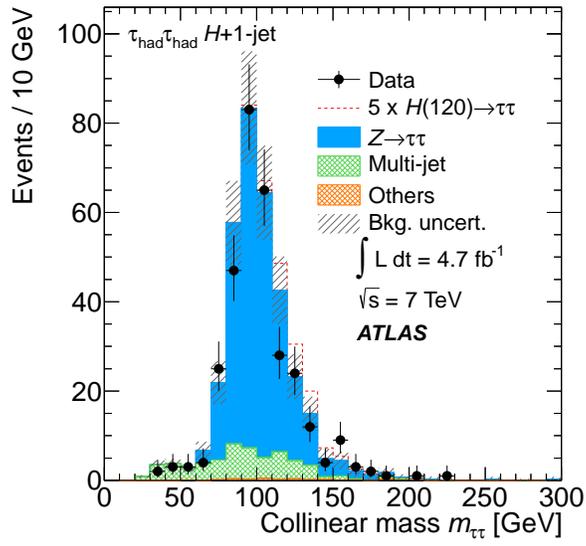} 
\end{center}
\vspace{-1em}
\caption{
Reconstructed  $m_{\tau\tau}$ of the selected events in the \Htthh\ channel.
Expectations from the Higgs boson signal ($m_H=120\,\gev$)  and from backgrounds are given. Results are shown after all selection criteria (see text).
For illustration only, the signal contribution has been scaled by a factor given in the legend.
}
\label{colmassSR}
\end{figure}

\clearpage
\section{Systematic uncertainties}
\label{sec:systematics}
Systematic uncertainties on the normalisation and shape
of the signal and background mass distributions are taken into account. 
These are treated either as fully correlated or uncorrelated across categories. In the case of partial correlations,
 the source is separated into correlated and uncorrelated components.
The dominant correlated systematic uncertainties are those on the measurement of the integrated luminosity and on the theoretical predictions of the signal production cross sections and decay 
branching ratios, as well as those related to detector response that impact the analyses through the reconstruction of electrons, muons, hadronic $\tau$ decays, jets, \met\ and $b$-tagging.

\paragraph{Theoretical uncertainties:} The Higgs boson cross section, branching ratios and their uncertainties are compiled in Refs.~\cite{LHCHiggsCrossSectionWorkingGroup:2011ti,LHCHiggsCrossSectionWorkingGroup:2012vm}.
The QCD scale uncertainties on the signal cross sections depend on $m_{H}$ and are of the order of 1\% for the VBF and $VH$ production modes
and in the range of 8--25\% for $gg\to H$ depending on jet multiplicity~\cite{Stewart:2011cf,ATL-PHYS-PUB-2011-011}. 
An uncertainty of 4--5\% is assumed for the inclusive cross section of the single vector boson and di-boson production mechanisms and
a relative uncertainty of 24\% is added in quadrature per additional jet.
For both \ttbar\ production 
and single top-quark production, the QCD scale uncertainties are in the range of 3--6\%~\cite{mochuwer1,beneke,kidonakis}.
The uncertainties related to the PDF amount to 8\% for the predominantly gluon-initiated processes, $gg\to H$ and \ttbar, 
and 4\% for the  predominantly quark-initiated processes, VBF, $VH$, single vector boson and di-boson production~\cite{Botje:2011sn,Lai:2010vv,Martin:2009iq,Ball:2011mu}. The systematic uncertainty arising from the choice of different sets of PDF is included.
In addition to the theoretical errors considered in Ref.~\cite{LHCHiggsCrossSectionWorkingGroup:2012vm}, other effects are taken into account. Uncertainties related to hadronisation effects are estimated by replacing PYTHIA with HERWIG. Effects due to initial and final 
state radiation are assessed with PYTHIA samples where the gluon emission is changed according to Ref.~\cite{Perugia}. The effect of a different choice of parton shower and underlying event parametrisation yields a total uncertainty of about 10\% on the acceptance of 
the Higgs boson produced via the VBF mechanism in the H+2jet VBF channel.

\paragraph{Detector-related uncertainties:}
The uncertainty on the integrated luminosity is considered as fully correlated across all analysis categories and amounts to 3.9\%~\cite{lumi,lumi2011}. The effect of pileup on the signal and background expectations is modelled in the Monte Carlo simulations and the corresponding uncertainty  is taken into account. 

Appropriate scale factors for the trigger efficiencies of electron, muon and hadronic $\tau$ triggers are obtained from data and applied to the Monte Carlo simulations. 
The associated systematic uncertainties are typically 1--2\%.
Differences between data and Monte Carlo simulations in the reconstruction and identification efficiencies of $\tau$ leptons, electrons and muons are taken into account, as well as the differences in the momentum scales and resolutions.

The systematic uncertainties on the hadronic $\tau$ decay identification efficiency are estimated from data samples enriched in $W\rightarrow\tau\nu$ and $\Ztotautau$ events and they are less than 4\%.
The energy scale uncertainties on the hadronic $\tau$ and jets are evaluated based on the single hadron response in the calorimeters~\cite{jes2010}.
In addition, the $\thad$ energy scale is validated in data samples enriched in $\Ztotautau$ events. The systematic uncertainties related to the $\tau$ and jet energy scale, 
resolution and $b$-veto are modelled as functions of  $\eta$ and $\pt$.
The jet and  $\tau$  energy scale  and resolution uncertainties are  treated as correlated and propagated to the \met\ calculation.
Uncertainties associated with the remaining pileup noise and cluster activity in the calorimeters are also considered as independent \met\ uncertainties.

The detector-related uncertainties depend on the event topology and are typically small compared to the theoretical uncertainties. 
The main exceptions are the jet energy scale uncertainty, which reaches up to $12\%$, and the  $\tau$ energy scale uncertainty, which is in the range 2--5\%.

\paragraph{Background modelling uncertainties:}
The modelling of the \Ztotautau\ background is performed with the data, as described in Section~\ref{sec:samples}. 
Corresponding uncertainties are obtained by propagating variations of the $\Ztomumu$ event selection and the muon energy subtraction procedure through the $\tau$-embedding procedure. 
Backgrounds with misreconstructed leptons and $\thad$ candidates are 
estimated with data and the uncertainty in the estimation lies in the range of 6--40\%. The uncertainty takes into account the dependence on the number of jets. The treatment of the other 
background processes varies across channels and the uncertainties related to the modelling are taken into account as described in Section~\ref{sec:backgrounds}.

\section{Statistical analysis}
\label{sec:statistics}

The statistical analysis of the data employs a binned likelihood function constructed as the product of the likelihood terms for each category.
A total of twelve categories are considered from the $\Httll$, $\Httlh$ and $\Htthh$ channels.
The likelihood in each category is a product over bins in the distributions of the  MMC mass, collinear mass or effective mass shown in Figs.~\ref{fig:tautaullmh}, \ref{fig:MMCMass_AllCategory} and \ref{colmassSR}.

The expected signal and background, as well as the observed number of events, 
in each bin of the mass distributions enter in the definition of the likelihood function $\mathcal{L}(\mu,\boldsymbol{\theta})$.
A ``signal strength'' parameter ($\mu$) multiplies the expected signal in each bin. The signal strength is a free parameter in the fit procedure. The value $\mu=0$ ($\mu=1$) corresponds to the absence (presence) of a Higgs boson signal with the SM production cross-section.
Signal and background predictions ($s$ and $b$) depend on systematic uncertainties that are
parametrised by nuisance parameters $\boldsymbol{\theta}$, which in turn are
constrained using Gaussian functions. The correlation of the systematic uncertainties across categories are taken into account:
\begin{equation}
\mathcal{L}(\mu, \theta) = 
\prod_{\textrm{bin}\,j} 
\text{Poisson}(N_j|\mu (s_j) + b_j)
\prod_{\theta} 
\text{Gaussian}(\theta|0,1).
\end{equation}
The expected signal and background event counts in each bin are functions of
$\boldsymbol{\theta}$. The parametrisation is chosen such that the rates in each
channel are log-normally distributed for a normally distributed $\boldsymbol{\theta}$.

The test statistic $q_\mu$ is defined as:
\begin{equation}
q_\mu = -2\ln\left(\mathcal{L}(\mu,\hat{\boldsymbol{\theta}}_\mu)/
                   \mathcal{L}(\hat{\mu},\hat{\boldsymbol{\theta}})\right),
\end{equation}
where $\hat{\mu}$ and $\hat{\boldsymbol{\theta}}$ refer to the global maximum of
the likelihood (with the constraint $0 \leq \hat{\mu} \leq \mu$) and
$\hat{\boldsymbol{\theta}}_\mu$ corresponds to the conditional maximum
likelihood of $\boldsymbol{\theta}$ for a given $\mu$.
This test statistic is used to compute exclusion limits following the modified
frequentist method known as CL$_{s}$~\cite{CLs_2002}.
The asymptotic approximation~\cite{Cowan:2010st} is used to evaluate the Gaussian probability 
density functions rather than performing pseudo-experiments and the procedure has been validated using ensemble tests.

The profile likelihood formalism used in this statistical analysis incorporates the information on the observed and expected number of events,
nuisance parameters, probability density functions and parameters of interest.
The statistical significance of an excess is evaluated in terms of the same
profile likelihood test statistic.
The expected sensitivity and the $\pm 1,2\,\sigma$ bands are obtained for the
background expectation in the absence of a Standard Model Higgs boson signal. The consistency with the background-only hypothesis is quantified using the $p$-value, the probability that the test statistic of a background-only experiment fluctuates to at least the observed one.

 \section{Results}
 \label{sec:results}

No significant excess is observed in the data compared to the SM expectation in any of the channels studied here. Exclusion limits at the 95\% confidence level, normalised to the Standard Model cross 
section times the branching ratio of $H\to\tau^+\tau^-$ ($\sigma_{\mathrm{SM}}$), are set as a function of the Higgs boson mass. Figure~\ref{fig:exclusion} shows expected and observed limits for the 
\begin{figure}[t!]
\begin{center}
\begin{tabular}{cc}
\includegraphics[width=0.45\textwidth]{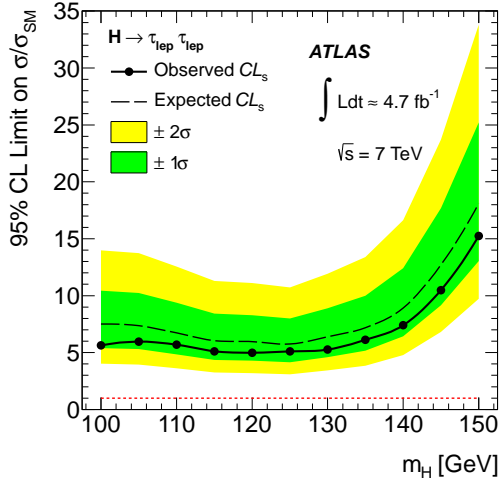}  &
\includegraphics[width=0.45\textwidth]{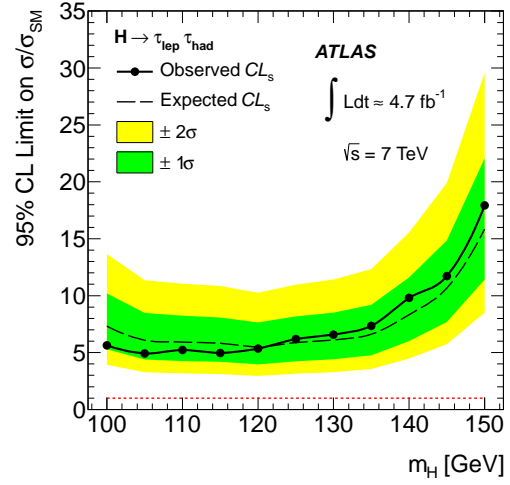} \\
(a) \Httll      & (b) \Httlh    \\
\includegraphics[width=0.45\textwidth]{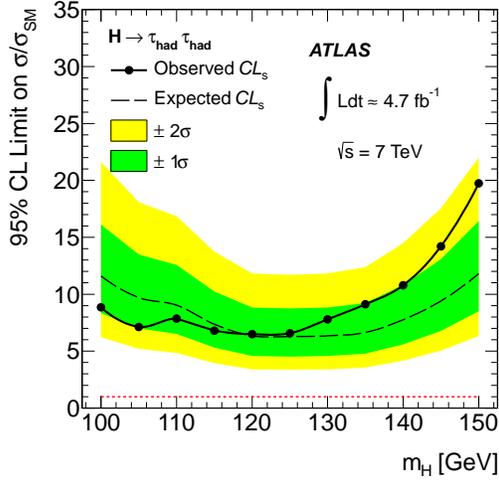} &
\includegraphics[width=0.45\textwidth]{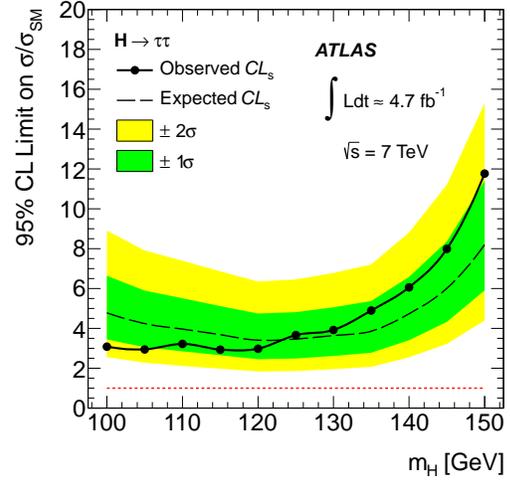}\\
(c)  \Htthh    & (d) Combined     \\
\end{tabular}
\caption[]{Observed (solid) and expected (dashed) 95\% confidence level upper limits on the Higgs boson cross section times branching ratio, 
normalised to the 
Standard Model expectation, as a function of the Higgs boson mass. Expected limits are given for the scenario with no signal. The bands around the dashed line indicate the expected statistical 
fluctuations of the limit. 
Results are given for the \Httll, \Httlh, and \Htthh\ channels independently and for all channels combined. \label{fig:exclusion}}
\end{center}
\end{figure}
individual channels and for the combined result. The combined expected limits vary between 3.4 and 8.2 times the predicted Standard Model cross section times branching ratio 
for the mass range 100--150\,\gev. The most sensitive categories are the $H+\!1$-jet category in the $\thad\thad$ channel, the $H+\!2$-jet VBF category in the $\tlep\thad$ channel and the $H+\!2$-jet 
VBF category in the $\tlep\tlep$ channel. The observed limits are in the range between 2.9 and 11.7 times the predicted Standard Model cross section times branching ratio 
for the same mass range. The most significant deviation from the background-only hypothesis is observed in the combined limit for $m_H=150\,\gev$ with a local $p$-value of 10\%, corresponding to a 
significance of 1.3~$\sigma$.

\section{Conclusions}

A search for a Higgs boson decaying in the $H\to\tau\tau$ channel has been performed 
with the ATLAS detector at the Large Hadron Collider. It uses the full 2011 data sample of 
$4.7\,\ifb$ collected at a centre-of-mass energy of 7 TeV. The \Httll, \Httlh and \Htthh\ 
decays are considered in this search. No 
significant excess is observed in the mass range of 100--150~\gev. The observed (expected)  
upper limits on the cross section times the branching ratio of $H\to\tau\tau$ are between 
2.9 (3.4) and 11.7 (8.2) times the SM prediction. These limits are similar to results recently 
reported by the CMS experiment~\cite{cmshiggstautaupaper}.

\clearpage
\acknowledgments{

We thank CERN for the very successful operation of the LHC, as well as the
support staff from our institutions without whom ATLAS could not be
operated efficiently.

We acknowledge the support of ANPCyT, Argentina; YerPhI, Armenia; ARC,
Australia; BMWF, Austria; ANAS, Azerbaijan; SSTC, Belarus; CNPq and FAPESP,
Brazil; NSERC, NRC and CFI, Canada; CERN; CONICYT, Chile; CAS, MOST and NSFC,
China; COLCIENCIAS, Colombia; MSMT CR, MPO CR and VSC CR, Czech Republic;
DNRF, DNSRC and Lundbeck Foundation, Denmark; EPLANET and ERC, European Union;
IN2P3-CNRS, CEA-DSM/IRFU, France; GNAS, Georgia; BMBF, DFG, HGF, MPG and AvH
Foundation, Germany; GSRT, Greece; ISF, MINERVA, GIF, DIP and Benoziyo Center,
Israel; INFN, Italy; MEXT and JSPS, Japan; CNRST, Morocco; FOM and NWO,
Netherlands; RCN, Norway; MNiSW, Poland; GRICES and FCT, Portugal; MERYS
(MECTS), Romania; MES of Russia and ROSATOM, Russian Federation; JINR; MSTD,
Serbia; MSSR, Slovakia; ARRS and MVZT, Slovenia; DST/NRF, South Africa;
MICINN, Spain; SRC and Wallenberg Foundation, Sweden; SER, SNSF and Cantons of
Bern and Geneva, Switzerland; NSC, Taiwan; TAEK, Turkey; STFC, the Royal
Society and Leverhulme Trust, United Kingdom; DOE and NSF, United States of
America.

The crucial computing support from all WLCG partners is acknowledged
gratefully, in particular from CERN and the ATLAS Tier-1 facilities at
TRIUMF (Canada), NDGF (Denmark, Norway, Sweden), CC-IN2P3 (France),
KIT/GridKA (Germany), INFN-CNAF (Italy), NL-T1 (Netherlands), PIC (Spain),
ASGC (Taiwan), RAL (UK) and BNL (USA) and in the Tier-2 facilities
worldwide.
}

\bibliographystyle{JHEP}
\bibliography{HtautauPaperJHEP}

\providecommand{\href}[2]{#2}\begingroup\raggedright\begin{thebibliography}{10}

\bibitem{prl_13_321}
{F.~Englert and R.~Brout},  {\em Phys. Rev. Lett.} {\bf 13} (1964) 321.

\bibitem{pl_12_132}
{P.~W.~Higgs},  {\em Phys. Lett.} {\bf 12} (1964) 132.

\bibitem{prl_13_508}
{P.~W.~Higgs},  {\em Phys. Rev. Lett.} {\bf 13} (1964) 508.

\bibitem{Guralnik:1964eu}
G.~S. Guralnik, C.~R. Hagen, and T.~W.~B. Kibble,  {\em Phys. Rev. Lett.} {\bf
  13} (1964) 585--587.

\bibitem{pr_145_1156}
{P.~W.~Higgs},  {\em Phys. Rev. D} {\bf 145} (1966) 1156.

\bibitem{Kibble:1967sv}
T.~W.~B. Kibble,  {\em Phys. Rev.} {\bf 155} (1967) 1554--1561.

\bibitem{lepew:2010vi}
{\textbf{ALEPH}, \textbf{DELPHI}, \textbf{L3}, \textbf{OPAL}, \textbf{SLD},
  \textbf{CDF}, and \textbf{D\O\ } Collaborations, The LEP Tevatron SLD
  Electroweak Working Group},  \href{http://xxx.lanl.gov/abs/1012.2367}{{\tt
  arXiv:1012.2367}}.

\bibitem{Barate:2003sz}
{\textbf{ALEPH}, \textbf{DELPHI}, \textbf{L3}, and \textbf{OPAL}
  Collaborations, The LEP Working Group for Higgs Boson Searches},  {\em Phys.
  Lett. B} {\bf 565} (2003) 61,
  [\href{http://xxx.lanl.gov/abs/hep-ex/0306033}{{\tt hep-ex/0306033}}].

\bibitem{TEVNPH:2012ab}
{\bf TEVNPH (Tevatron New Phenomena and Higgs Working Group), CDF and D0}
  Collaboration, \href{http://xxx.lanl.gov/abs/1203.3774}{{\tt
  arXiv:1203.3774}}. Preliminary results prepared for the Winter 2012
  Conferences.

\bibitem{atlashiggspaper}
{\bf {ATLAS}} Collaboration, {\em Phys. Lett. B} {\bf 710} (2012) 49,
  [\href{http://xxx.lanl.gov/abs/1202.1408}{{\tt arXiv:1202.1408}}].

\bibitem{Chatrchyan:2012tx}
{\bf CMS} Collaboration, {\em Phys. Lett. B} {\bf 710} (2012) 26,
  [\href{http://xxx.lanl.gov/abs/1202.1488}{{\tt arXiv:1202.1488}}].

\bibitem{atlas-det}
{\bf ATLAS} Collaboration, {\em JINST} {\bf 3} (2008) S08003.

\bibitem{Djouadi1991}
A.~Djouadi, M.~Spira, and P.~M. Zerwas,  {\em Phys. Lett. B} {\bf 264} (1991)
  440--446.

\bibitem{Dawson1991}
S.~Dawson,  {\em Nucl. Phys. B} {\bf 359} (1991) 283--300.

\bibitem{Spira1995}
M.~Spira, A.~Djouadi, D.~Graudenz, and P.~M. Zerwas,  {\em Nucl. Phys. B} {\bf
  453} (1995) 17--82, [\href{http://xxx.lanl.gov/abs/hep-ph/9504378}{{\tt
  hep-ph/9504378}}].

\bibitem{Harlander:2002wh}
{R.~Harlander and W.~B.~Kilgore},  {\em Phys. Rev. Lett.} {\bf 88} (2002)
  201801, [\href{http://xxx.lanl.gov/abs/hep-ph/0201206}{{\tt
  hep-ph/0201206}}].

\bibitem{Anastasiou:2002yz}
C.~Anastasiou and K.~Melnikov,  {\em Nucl. Phys. B} {\bf 646} (2002) 220,
  [\href{http://xxx.lanl.gov/abs/hep-ph/0207004}{{\tt hep-ph/0207004}}].

\bibitem{Ravindran:2003um}
{V.~Ravindran, J.~Smith, W.~L.~van Neerven},  {\em Nucl. Phys. B} {\bf 665}
  (2003) 325, [\href{http://xxx.lanl.gov/abs/hep-ph/0302135}{{\tt
  hep-ph/0302135}}].

\bibitem{Catani:2003zt}
S.~Catani, D.~de~Florian, M.~Grazzini, and P.~Nason,  {\em JHEP} {\bf 0307}
  (2003) 028, [\href{http://xxx.lanl.gov/abs/hep-ph/0306211}{{\tt
  hep-ph/0306211}}].

\bibitem{Aglietti:2004nj}
{U.~Aglietti, R.~Bonciani, G.~Degrassi, and A.~Vicini},  {\em Phys. Lett. B}
  {\bf 595} (2004) 432, [\href{http://xxx.lanl.gov/abs/hep-ph/0404071}{{\tt
  hep-ph/0404071}}].

\bibitem{Actis:2008ug}
S.~Actis, G.~Passarino, C.~Sturm, and S.~Uccirati,  {\em Phys. Lett. B} {\bf
  670} (2008) 12--17, [\href{http://xxx.lanl.gov/abs/0809.1301}{{\tt
  arXiv:0809.1301}}].

\bibitem{Anastasiou:2008tj}
{C.~Anastasiou, R.~Boughezal, and F.~Petriello},  {\em JHEP} {\bf 0904} (2009)
  003, [\href{http://xxx.lanl.gov/abs/0811.3458}{{\tt arXiv:0811.3458}}].

\bibitem{deFlorian:2009hc}
D.~de~Florian and M.~Grazzini,  {\em Phys. Lett. B} {\bf 674} (2009) 291,
  [\href{http://xxx.lanl.gov/abs/0901.2427}{{\tt arXiv:0901.2427}}].

\bibitem{Baglio:2010ae}
{J.~Baglio and A.~Djouadi},  {\em JHEP} {\bf 1103} (2011) 055,
  [\href{http://xxx.lanl.gov/abs/1012.0530}{{\tt arXiv:1012.0530}}].

\bibitem{Ciccolini:2007jr}
M.~Ciccolini, A.~Denner, and S.~Dittmaier,  {\em Phys. Rev. Lett.} {\bf 99}
  (2007) 161803, [\href{http://xxx.lanl.gov/abs/0707.0381}{{\tt
  arXiv:0707.0381}}].

\bibitem{Ciccolini:2007ec}
M.~Ciccolini, A.~Denner, and S.~Dittmaier,  {\em Phys. Rev. D} {\bf 77} (2008)
  013002, [\href{http://xxx.lanl.gov/abs/0710.4749}{{\tt arXiv:0710.4749}}].

\bibitem{Arnold:2008rz}
K.~Arnold, M.~Bahr, G.~Bozzi, F.~Campanario, C.~Englert, {\em et.~al.},  {\em
  Comput.Phys.Commun.} {\bf 180} (2009) 1661--1670,
  [\href{http://xxx.lanl.gov/abs/0811.4559}{{\tt arXiv:0811.4559}}].

\bibitem{Bolzoni:2010xr}
{P.~Bolzoni, F.~Maltoni, S.-O.~Moch, and M.~Zaro},  {\em Phys. Rev. Lett.} {\bf
  105} (2010) 011801, [\href{http://xxx.lanl.gov/abs/1003.4451}{{\tt
  arXiv:1003.4451}}].

\bibitem{Han:1991ia}
T.~Han and S.~Willenbrock,  {\em Phys. Lett. B} {\bf 273} (1991) 167--172.

\bibitem{Brein:2003wg}
{O.~Brein, A.~Djouadi, and R.~Harlander},  {\em Phys. Lett. B} {\bf 579} (2004)
  149, [\href{http://xxx.lanl.gov/abs/hep-ph/0307206}{{\tt hep-ph/0307206}}].

\bibitem{Ciccolini:2003jy}
{M.~L.~Ciccolini, S.~Dittmaier, and M.~Kramer},  {\em Phys. Rev. D} {\bf 68}
  (2003) 073003, [\href{http://xxx.lanl.gov/abs/hep-ph/0306234}{{\tt
  hep-ph/0306234}}].

\bibitem{Djouadi:1997yw}
{A.~Djouadi, J.~Kalinowski, and M.~Spira},  {\em Comput. Phys. Commun.} {\bf
  108} (1998) 56, [\href{http://xxx.lanl.gov/abs/hep-ph/9704448}{{\tt
  hep-ph/9704448}}].

\bibitem{Alioli:2008tz}
{S. Alioli, P. Nason, and C. Oleari, and E. Re},  {\em JHEP} {\bf 0904} (2009)
  002, [\href{http://xxx.lanl.gov/abs/0812.0578}{{\tt arXiv:0812.0578}}].

\bibitem{Nason:2009ai}
P.~Nason and C.~Oleari,  {\em JHEP} {\bf 1002} (2010) 037,
  [\href{http://xxx.lanl.gov/abs/0911.5299}{{\tt arXiv:0911.5299}}].

\bibitem{pythia}
{T.~Sjostrand, S.~Mrenna, and P.~Z.~Skands},  {\em JHEP} {\bf 0605} (2006) 026.

\bibitem{deFlorian:2011xf}
{D.~de~Florian et al.},  {\em JHEP} {\bf 11} (2011) 064,
  [\href{http://xxx.lanl.gov/abs/1109.2109}{{\tt arXiv:1109.2109}}]. For Higgs
  boson $p_{\mathrm{T}} > m_{H}$, the calculation is switched from NLO+NLL to
  NLO.

\bibitem{alpgen}
M.~L. Mangano {\em et.~al.},  {\em JHEP} {\bf 0307} (2003) 001.

\bibitem{herwig}
G.~Corcella {\em et.~al.},  {\em JHEP} {\bf 0101} (2001) 010.

\bibitem{alwall-2008-53}
{J.~ Alwall et al.},  {\em Eur. Phys. J. C} {\bf 53} (2008) 473.

\bibitem{mcatnlo}
S.~Frixione and B.~R. Webber,  {\em JHEP} {\bf 06} (2002) 029,
  [\href{http://xxx.lanl.gov/abs/hep-ph/0204244}{{\tt hep-ph/0204244}}].

\bibitem{jimmy}
J.~M. Butterworth, J.~R. Forshaw, and M.~H. Seymour,  {\em Z. Phys. C} {\bf 72}
  (1996) 637--646, [\href{http://xxx.lanl.gov/abs/hep-ph/9601371}{{\tt
  hep-ph/9601371}}].

\bibitem{Kersevan:2004yg}
{B.~P.~Kersevan and E.~Richter-Was},
  \href{http://xxx.lanl.gov/abs/hep-ph/0405247}{{\tt hep-ph/0405247}}.

\bibitem{Jadach:1993hs}
S.~Jadach, Z.~Was, R.~Decker, and J.~H. Kuhn,  {\em Comput. Phys. Commun.} {\bf
  76} (1993) 361--380.

\bibitem{Golonka:2005pn}
P.~Golonka and Z.~Was,  {\em Eur. Phys. J. C} {\bf 45} (2006) 97--107,
  [\href{http://xxx.lanl.gov/abs/hep-ph/0506026}{{\tt hep-ph/0506026}}].

\bibitem{ct10}
H.-L. Lai {\em et.~al.},  {\em Phys. Rev. D} {\bf 82} (2010) 074024,
  [\href{http://xxx.lanl.gov/abs/1007.2241}{{\tt arXiv:1007.2241}}].

\bibitem{Pumplin:2002vw}
J.~Pumplin {\em et.~al.},  {\em JHEP} {\bf 07} (2002) 012,
  [\href{http://xxx.lanl.gov/abs/hep-ph/0201195}{{\tt hep-ph/0201195}}].

\bibitem{mrst}
A.~Sherstnev and R.~S. Thorne,  {\em Eur. Phys. J C} {\bf 55} (2009) 553.

\bibitem{GEANT4}
{S.~Agostinelli et al.},  {\em Nucl. Instrum. Meth. A} {\bf 506} (2003) 250.

\bibitem{atlassim}
{\bf ATLAS} Collaboration, {\em Eur. Phys. J. C} {\bf 70} (2010) 823--874,
  [\href{http://xxx.lanl.gov/abs/1005.4568}{{\tt arXiv:1005.4568}}].

\bibitem{egammapaper}
{\bf ATLAS} Collaboration, {\em Eur.Phys.J. C} {\bf 72} (2012) 1909,
  [\href{http://xxx.lanl.gov/abs/1110.3174}{{\tt arXiv:1110.3174}}].

\bibitem{ATLASWZFirstConfNote}
{\bf ATLAS} Collaboration, {\em JHEP} {\bf 1012} (2010) 060,
  [\href{http://xxx.lanl.gov/abs/1010.2130}{{\tt arXiv:1010.2130}}].

\bibitem{AntiKT}
M.~Cacciari, G.~P. Salam, and G.~Soyez,  {\em JHEP} {\bf 0804} (2008) 063,
  [\href{http://xxx.lanl.gov/abs/0802.1189}{{\tt arXiv:0802.1189}}].

\bibitem{ATLAS-btag-algs}
{\bf {ATLAS}} Collaboration, ATLAS-CONF-2011-102,
  [http://cdsweb.cern.ch/record/1369219].

\bibitem{ATLAS-btag-eff}
{\bf {ATLAS}} Collaboration, ATLAS-CONF-2012-043,
  [http://cdsweb.cern.ch/record/1435197].

\bibitem{ATLAS-btag-mis}
{\bf {ATLAS}} Collaboration, ATLAS-CONF-2012-040,
  [http://cdsweb.cern.ch/record/1435194].

\bibitem{TauID}
{\bf ATLAS} Collaboration, ATLAS-CONF-2011-152,
  [http://cdsweb.cern.ch/record/1398195].

\bibitem{ATLASmetNEW}
{\bf ATLAS} Collaboration, {\em Eur. Phys. J. C} {\bf 72} (2012) 1844,
  [\href{http://xxx.lanl.gov/abs/1108.5602}{{\tt arXiv:1108.5602}}].

\bibitem{atlastrigger}
{\bf ATLAS} Collaboration, {\em Eur.Phys.J.} {\bf C72} (2012) 1849,
  [\href{http://xxx.lanl.gov/abs/1110.1530}{{\tt arXiv:1110.1530}}].

\bibitem{Rainwater:1998kj}
D.~L. Rainwater, D.~Zeppenfeld, and K.~Hagiwara,  {\em Phys. Rev. D} {\bf 59}
  (1998) 014037, [\href{http://xxx.lanl.gov/abs/hep-ph/9808468}{{\tt
  hep-ph/9808468}}].

\bibitem{Plehn:1999xi}
{T.~Plehn, D.~L.~Rainwater and D.~Zeppenfeld},  {\em Phys. Rev. D} {\bf 61}
  (2000) 093005, [\href{http://xxx.lanl.gov/abs/hep-ph/9911385}{{\tt
  hep-ph/9911385}}].

\bibitem{Asai:2004ws}
S.~Asai {\em et.~al.},  {\em Eur. Phys. J. C} {\bf 32S2} (2004) 19--54,
  [\href{http://xxx.lanl.gov/abs/hep-ph/0402254}{{\tt hep-ph/0402254}}].

\bibitem{Mellado:2004tj}
B.~Mellado, W.~Quayle, and S.~L. Wu,  {\em Phys. Lett. B} {\bf 611} (2005)
  60--65, [\href{http://xxx.lanl.gov/abs/hep-ph/0406095}{{\tt
  hep-ph/0406095}}].

\bibitem{cll2}
{R.K. Ellis, I. Hinchliffe, M. Soldate and J.J. Van der Bij},  {\em Nucl. Phys.
  B} {\bf 297} (1988) 221.

\bibitem{Elagin:2010aw}
A.~Elagin, P.~Murat, A.~Pranko, and A.~Safonov,  {\em Nucl. Instrum. Meth. A}
  {\bf 654} (2011) 481--489, [\href{http://xxx.lanl.gov/abs/1012.4686}{{\tt
  arXiv:1012.4686}}].

\bibitem{LHCHiggsCrossSectionWorkingGroup:2011ti}
{S.~Dittmaier et al. (LHC Higgs Cross Section Working Group)},
  \href{http://xxx.lanl.gov/abs/1101.0593}{{\tt arXiv:1101.0593}}.

\bibitem{LHCHiggsCrossSectionWorkingGroup:2012vm}
{S.~Dittmaier et al. (LHC Higgs Cross Section Working Group)},
  \href{http://xxx.lanl.gov/abs/1201.3084}{{\tt arXiv:1201.3084}}.

\bibitem{Stewart:2011cf}
I.~W. Stewart and F.~J. Tackmann,  {\em Phys. Rev. D} {\bf 85} (2012) 034011,
  [\href{http://xxx.lanl.gov/abs/1107.2117}{{\tt arXiv:1107.2117}}].

\bibitem{ATL-PHYS-PUB-2011-011}
{\textbf{ATLAS} and \textbf{CMS} Collaborations},  ATL-PHYS-PUB-2011-011,
  CMS-NOTE-2011-005.

\bibitem{mochuwer1}
{S. Moch and P. Uwer},  {\em Phys. Rev. D} {\bf 78} (2008) 034003.

\bibitem{beneke}
{M. Beneke et al.},  {\em Phys. Lett. B} {\bf 690} (2010) 483.

\bibitem{kidonakis}
{N. Kidonakis},  {\em Phys. Rev. D} {\bf 83} (2011) 091503,
  [\href{http://xxx.lanl.gov/abs/1103.2792}{{\tt arXiv:1103.2792}}].

\bibitem{Botje:2011sn}
{M.~Botje et al.},  \href{http://xxx.lanl.gov/abs/1101.0538}{{\tt
  arXiv:1101.0538}}.

\bibitem{Lai:2010vv}
{H.-L. Lai et al.},  {\em Phys. Rev. D} {\bf 82} (2010)
  [\href{http://xxx.lanl.gov/abs/1007.2241}{{\tt arXiv:1007.2241}}].

\bibitem{Martin:2009iq}
A.~D. Martin, W.~J. Stirling, R.~S. Thorne, and G.~Watt,  {\em Eur. Phys. J. C}
  {\bf 63} (2009) 189--285, [\href{http://xxx.lanl.gov/abs/0901.0002}{{\tt
  arXiv:0901.0002}}].

\bibitem{Ball:2011mu}
{R. D. Ball et al.},  {\em Nucl. Phys. B} {\bf 849} (2011)
  [\href{http://xxx.lanl.gov/abs/1101.1300}{{\tt arXiv:1101.1300}}].

\bibitem{Perugia}
{P.~Z.~Skands},  {\em Phys. Rev. D} {\bf 82} (2010) 074018,
  [\href{http://xxx.lanl.gov/abs/1005.3457}{{\tt arXiv:1005.3457}}].

\bibitem{lumi}
{\bf ATLAS} Collaboration, {\em Eur. Phys. J. C} {\bf 71} (2011) 1630,
  [\href{http://xxx.lanl.gov/abs/1101.2185}{{\tt arXiv:1101.2185}}].

\bibitem{lumi2011}
{\bf ATLAS} Collaboration, ATLAS-CONF-2011-116
  [http://cdsweb.cern.ch/record/1376384].

\bibitem{jes2010}
{\bf ATLAS} Collaboration, \href{http://xxx.lanl.gov/abs/1112.6426}{{\tt
  arXiv:1112.6426}}. Submitted to EPJC.

\bibitem{CLs_2002}
{A.L.~Read},  {\em J. Phys. G} {\bf 28} (2002) 2693.

\bibitem{Cowan:2010st}
{G.~Cowan, K.~Cranmer, E.~Gross, and O.~Vitells},  {\em Eur. Phys. J. C} {\bf
  71} (2011) 1554.

\bibitem{cmshiggstautaupaper}
{\bf {CMS}} Collaboration, {\em Phys. Lett. B} {\bf 713} (2012) 68,
  [\href{http://xxx.lanl.gov/abs/1202.4083}{{\tt arXiv:1202.4083}}].

\end{thebibliography}\endgroup
\clearpage
\begin{flushleft}
{\Large The ATLAS Collaboration}

\bigskip

G.~Aad$^{\rm 48}$,
B.~Abbott$^{\rm 111}$,
J.~Abdallah$^{\rm 11}$,
S.~Abdel~Khalek$^{\rm 115}$,
A.A.~Abdelalim$^{\rm 49}$,
O.~Abdinov$^{\rm 10}$,
B.~Abi$^{\rm 112}$,
M.~Abolins$^{\rm 88}$,
O.S.~AbouZeid$^{\rm 158}$,
H.~Abramowicz$^{\rm 153}$
% ...
\bigskip

% ...
$^{10}$ Institute of Physics, Azerbaijan Academy of Sciences, Baku, Azerbaijan\\
$^{11}$ Institut de F\'isica d'Altes Energies and Departament de F\'isica de la Universitat Aut\`onoma  de Barcelona and ICREA, Barcelona, Spain\\
% ...
$^{48}$ Fakult\"{a}t f\"{u}r Mathematik und Physik, Albert-Ludwigs-Universit\"{a}t, Freiburg i.Br., Germany\\
$^{49}$ Section de Physique, Universit\'e de Gen\`eve, Geneva, Switzerland\\
% ...
$^{88}$ Department of Physics and Astronomy, Michigan State University, East Lansing MI, United States of America\\
% ...
$^{111}$ Homer L. Dodge Department of Physics and Astronomy, University of Oklahoma, Norman OK, United States of America\\
$^{112}$ Department of Physics, Oklahoma State University, Stillwater OK, United States of America\\
% ...
$^{115}$ LAL, Universit\'e Paris-Sud and CNRS/IN2P3, Orsay, France\\
% ...
$^{153}$ Raymond and Beverly Sackler School of Physics and Astronomy, Tel Aviv University, Tel Aviv, Israel\\
% ...
$^{158}$ Department of Physics, University of Toronto, Toronto ON, Canada\\
% ...
\end{flushleft}

\end{document}
\bye